\documentclass[
reprint,
longbibliography,
aps,
twoside,
superscriptaddress,
bibnotes,
floatfix
]{revtex4-2}

\usepackage{physics}
\usepackage{dsfont}
\usepackage{color,newfloat}
\usepackage{graphicx}
\usepackage{dcolumn}
\usepackage{bm}
\usepackage{amsmath,amssymb,amsthm,amsfonts}
\usepackage{mathtools}
\usepackage{multirow}
\usepackage{hyperref}
\usepackage[all]{hypcap}
\usepackage{chngcntr}
\usepackage{enumerate}

\newcommand{\TR}[1]{\mathrm{Tr}\left(#1\right)}

\newcommand{\vdd}[2]{\bm{r}_{#1}\cdot\bm{r}_{#2}}

\begin{document}

\title{Distinguishability and mixedness in quantum interference}

\author{Alex E Jones}
\email{these authors contributed equally: \\a.jones@bristol.ac.uk\\shreya.kumar@fmq.uni-stuttgart.de}
\affiliation{Quantum Engineering Technology Labs, H. H. Wills Physics Laboratory and Department of Electrical and Electronic Engineering, University of Bristol, Bristol BS8 1FD, UK}
\author{Shreya Kumar}
\email{these authors contributed equally: \\a.jones@bristol.ac.uk\\shreya.kumar@fmq.uni-stuttgart.de}
\affiliation{Institute for Functional Matter and Quantum Technologies, University of Stuttgart, 70569 Stuttgart, Germany}
\affiliation{Center for Integrated Quantum Science and Technology (IQST), University of Stuttgart, 70569 Stuttgart, Germany}
\author{Simone D'Aurelio}
\affiliation{Institute for Functional Matter and Quantum Technologies, University of Stuttgart, 70569 Stuttgart, Germany}
\affiliation{Center for Integrated Quantum Science and Technology (IQST), University of Stuttgart, 70569 Stuttgart, Germany}
\author{Matthias Bayerbach}
\affiliation{Institute for Functional Matter and Quantum Technologies, University of Stuttgart, 70569 Stuttgart, Germany}
\affiliation{Center for Integrated Quantum Science and Technology (IQST), University of Stuttgart, 70569 Stuttgart, Germany}
\author{Adrian J Menssen}
\affiliation{Research Laboratory of Electronics, Massachusetts Institute of Technology, Cambridge, Massachusetts 02139, USA}
\author{Stefanie Barz}
\email{stefanie.barz@fmq.uni-stuttgart.de}
\affiliation{Institute for Functional Matter and Quantum Technologies, University of Stuttgart, 70569 Stuttgart, Germany}
\affiliation{Center for Integrated Quantum Science and Technology (IQST), University of Stuttgart, 70569 Stuttgart, Germany}


\begin{abstract}
We study the impact of distinguishability and mixedness -- two fundamental properties of quantum states -- on quantum interference. We show that these can influence the interference of multiple particles in different ways, leading to effects that cannot be observed in the interference of two particles alone. This is demonstrated experimentally by interfering three independent photons in pure and mixed states and observing their different multiphoton interference, despite exhibiting the same two-photon Hong-Ou-Mandel (HOM) interference. Besides its fundamental relevance, our observation has important implications for quantum technologies relying on photon interference.
\end{abstract}

\maketitle


\section{Introduction}
Quantum interference is a defining feature of quantum physics, leading to behaviour that puts it in sharp contrast to its classical counterpart. Beyond its importance at a fundamental level, interference is also a crucial component for many modern quantum technologies. For example, those relying on photon interference include photonic quantum computing, optical demonstrations of quantum advantage, quantum metrology and quantum networks~\cite{OBrien2009,Kok2007,Zhong2020,Polino2020, Sangouard2011,Kimble2008,Pompili2021}. However, quantum interference suffers from an intrinsic trade-off between which-path information and interference strength~\cite{Englert1996, Dittel2021}. Therefore, a key requirement for high-quality operation of these technologies is that the photons are indistinguishable and have a high purity. Otherwise the quantum interference degrades and this leads to a reduction in the fidelity of quantum operations.

Extensive theoretical and experimental work has been developed for analysing the effect of photon distinguishability in computationally demanding tasks like boson sampling~\cite{Tichy2015,Tillmann2015,Shchesnovich2015,Shchesnovich2018,Renema2018,Menssen2017,Jones2020, Minke2021,Brod2019,Viggianiello2018}. A common approach to quantifying similarity of photons is to perform a Hong-Ou-Mandel (HOM) interference test between pairs of them. One photon is injected into each input of a balanced beam splitter and the relative time delay is varied over the coherence time of the photons' wavepackets~\cite{Hong1987}. The visibility of the variation in coincidence counts gives a measure of the overlap of the photons' wavefunctions, and so can characterise the similarity and quality of photons~\cite{Bouchard2021}. Besides distinguishability, another important photon error is mixedness. This implies some degree of entanglement with the environment and also leads to decreased interference strength. Existing theoretical work captures the effects of both of these errors but qualitative differences between the resulting behaviour are unclear~\cite{Tichy2015,Shchesnovich2015,Rohde2015}.

Here, we show for the first time that discrimination between distinguishability and purity in the interference of independent photons requires at least three photons. We demonstrate this experimentally by preparing sets of three photons in pure and mixed states and interfering them in a multiport splitter. We use intuitive geometric interpretations to show that in both cases the HOM visibilities are the same, but other multiphoton output statistics differ. Our result underscores that two-particle interferences are generally insufficient for predicting the nature of interference in larger systems of mixed quantum states.

\section{Distinguishable and mixed states}
When we talk about photons' states we will make a distinction between the \textit{external} degrees of freedom resolved by detectors -- in our case, spatial position -- and the \textit{internal} degrees of freedom that are not resolved -- such as colour or polarisation.
We will investigate distinguishability in interference by starting with a set of independent photons, prepared with one photon per external state, and internal states that determine their distinguishability. 
The photons then evolve through an interferometer that evolves external states but leaves internal states unchanged. Finally the photons are detected at the outputs. 

\begin{figure}[hb]
	\centering
	\includegraphics[width=0.44\textwidth]{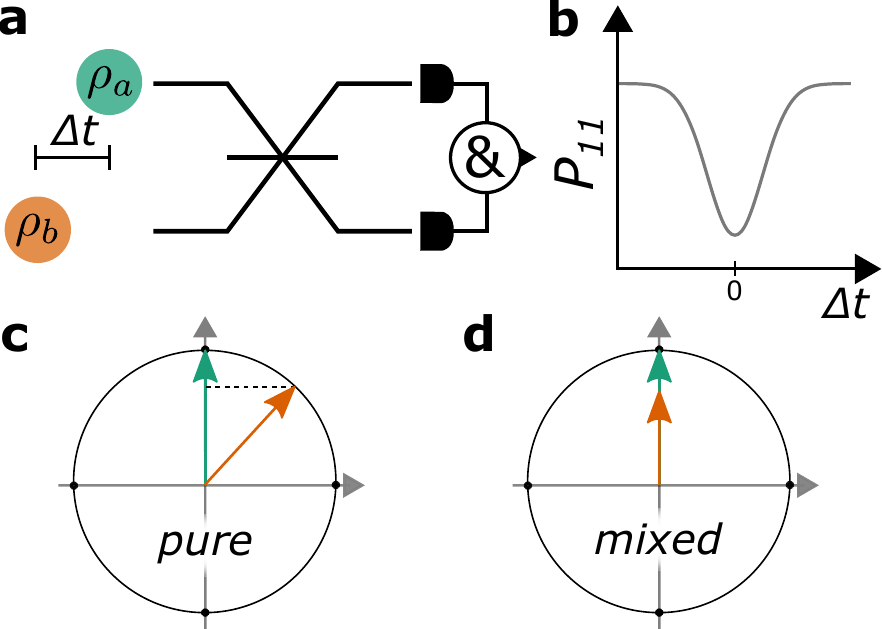}
	\caption{\label{fig:fig1}\textbf{a} HOM dip measurement where photons with internal states $\rho_{a}$ and $\rho_{b}$ and relative time delay $\Delta t$ interfere on a balanced beam splitter. \textbf{b} We here assume the photons occupy identical Gaussian wavepackets. The output coincidence probability $P_{11}$ reaches a minimum at zero relative delay and the visibility depends on the overlap of internal states. For a two-dimensional internal space, this overlap of Bloch vectors is reduced by distinguishability (\textbf{c}) and mixedness (\textbf{d}) in the same way, giving identical HOM dip visibilities.}
\end{figure}

If photons differ in their internal states then there is, in principle, information that can distinguish between the photon paths in an interferometer leading to a particular output pattern~\cite{Tillmann2015,Tichy2015,Shchesnovich2015}. For example, in HOM interference the relative arrival time of two photons at a beam splitter allows discrimination between the two paths leading to output coincidence detection -- namely both photons being transmitted or both being reflected. As the relative time delay approaches zero, the paths become indistinguishable. This strengthens the destructive interference that suppresses output coincidences and leads to a characteristic HOM dip (see Fig.~\ref{fig:fig1} a, b).

If the two photons instead have pure internal states $\ket{a}$ and $\ket{b}$ that are different in some other degree of freedom besides arrival time, then the strength of coincidence suppression at zero relative delay is determined by the pairwise distinguishability $\abs{\braket{a}{b}}^2$. This can be measured using the HOM dip visibility, $\mathcal{V}=\rm{(max-min)/max}$. Extending to more simultaneous photons and a larger interferometer, the introduction of another photon with state $\ket{c}$ means there are four distinguishing parameters: three pairwise distinguishabilities and the appearance of a multiparticle phase $\varphi_{abc}=\mathrm{arg}\left(\braket{a}{b}\braket{b}{c}\braket{c}{a}\right)$~\cite{Menssen2017}. Adding more photons leads to more pairwise distinguishabilities and additional multiparticle phases with a similar form. For $N$ pure photons, with one photon per external state, at most $(N-1)^2$ real parameters describe their distinguishability~\cite{Shchesnovich2018, Jones2020}.

This reasoning applies to pure states only. More generally, quantum systems such as photons can be in mixed states. For example, heralding from a spectrally entangled pair source leads to mixed photon states. The similarity of two photons in states $\rho_{a}$ and $\rho_{b}$ is then described by the real pairwise trace $\mathrm{Tr}\left(\rho_{a}\rho_{b}\right)$. Extending to more photons, similarity is related to traces of products of their density matrices and mixedness means the largest number of distinguishing parameters for $N$ coincident photons increases to $(N!-1)$~\cite{Shchesnovich2015,Stanisic2018}. Constructing pure state decompositions for the mixed states involved means that interference can be expressed as an incoherent sum of pure state interferences~\cite{Tichy2015}. However, such a presentation obscures intuitive aspects of interference that can be captured through careful consideration of the distinguishing parameters. In this paper we investigate the form of these parameters for two and three photons and show how distinguishability and mixedness can affect multiphoton interference differently.

\section{Two-photon interference}
We begin by showing that two-photon interference does not discriminate between distinguishability and mixedness. Consider two photons with states $\rho_{a}$ and $\rho_{b}$ impinging on a balanced beam splitter. If the probability of observing output coincidences $P_{11}$ is monitored as the relative time delay is varied, a HOM dip is observed (Fig.~\ref{fig:fig1}a, b). Its visibility is determined by the pairwise trace $\mathrm{Tr}\left(\rho_{a}\rho_{b}\right)$. This quantity can be interpreted geometrically by considering an internal space with two dimensions -- here the minimum number needed for two photons to exhibit complete distinguishability.

Let Bloch vectors $\bm{r}_{a}$ and $\bm{r}_{b}$ describe the internal states $\rho_{a}$ and $\rho_{b}$ respectively, so $\bm{r}_{j}=\mathrm{Tr}\left(\rho_{j}\boldsymbol{\sigma}\right)$, $j=a,b$, where $\boldsymbol{\sigma}$ is the vector of Pauli matrices. The pairwise trace is related to the dot product as
\begin{equation}\label{eqn:TrAB}
\TR{\rho_{a}\rho_{b}}=\frac{1}{2}\left(1+ \bm{r}_{a} \cdot \bm{r}_{b} \right).
\end{equation}
The lengths and relative orientation of the Bloch vectors describe the purities and distinguishability of the states, respectively, and affect the pairwise trace in the same way (see Fig.~\ref{fig:fig1}c, d). This also holds for two vectors describing internal states in a higher-dimensional space, where the dot product determines the pairwise trace. Thus, the visibility of HOM interference between two independent photons cannot discriminate between distinguishability and mixedness.

\section{Three-photon interference}
The interference of three photons depends on five distinguishing parameters: three real pairwise traces and also the generally \textit{complex} triple trace $\TR{\rho_{a}\rho_{b}\rho_{c}}$. Here we consider qubit internal states that permit an intuitive geometric description of the triple trace. Associating a Bloch vector $\bm{r}_{j}$ to each state $\rho_{j}, j=a,b,c$, we find
\begin{equation}\label{eqn:Trabc}
\TR{\rho_{a}\rho_{b}\rho_{c}}=\frac{1}{4}\left(1+\vdd{a}{b}+\vdd{a}{c}+\vdd{b}{c}+iV_{abc}\right).
\end{equation}
The dot products describe pairwise similarities and the imaginary component encodes a collective description via the scalar triple product of the three Bloch vectors: $V_{abc}=\bm{r}_{a}\cdot(\bm{r}_{b}\times\bm{r}_{c})$. This has a magnitude given by the volume of the corresponding parallelepiped and a sign set by its orientation in a right-handed frame (see Fig.~\ref{fig:fig2}a). It is also anti-symmetric under pairwise swaps of the vectors. If the Bloch vectors are coplanar then $V_{abc}=0$ and the triple trace is real and fully determined by pairwise traces. Otherwise it contains information not captured by dot products alone.

\section{Pure and mixed preparations of three photons}
We now consider two preparations of three photons: one where pairwise similarities of pure states are governed by distinguishability, and another where the pairwise similarities are determined by state purity. HOM visibilities for the two preparations could be the same, but we will see that $V_{abc}$ plays an important role in three-photon interference.

\begin{figure}[t]
	\centering
	\includegraphics[width=0.475\textwidth]{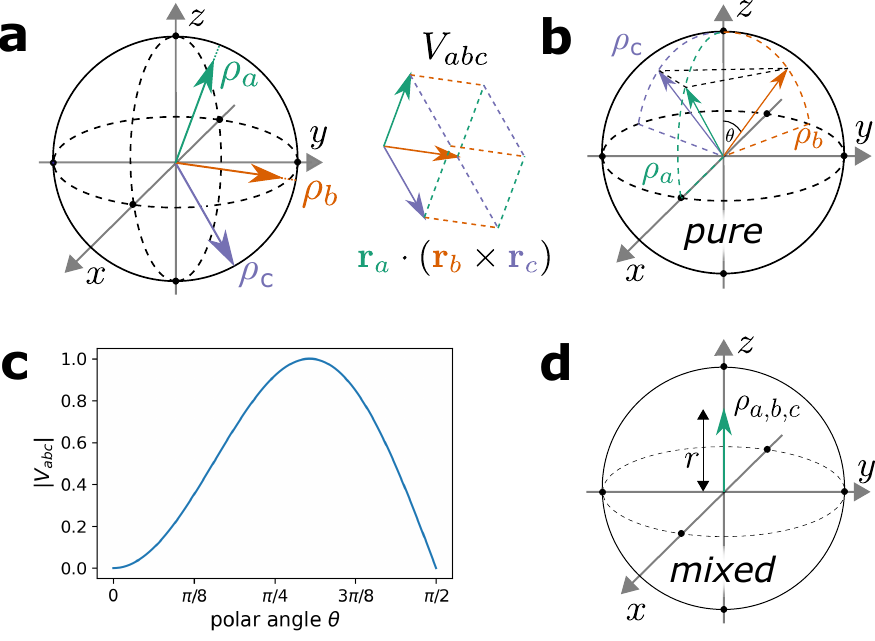}
	\caption{\label{fig:fig2}\textbf{a} The scalar triple product of three Bloch vectors $V_{abc}$ encodes a collective description of the states. \textbf{b} Pure qubit internal states equally spaced in azimuthal angle but with varying $\theta$ (Eqn.~\ref{eqn:prep1}). \textbf{c} Magnitude of the volume $V_{abc}$ for the pure state preparation. \textbf{d} Mixed states where pairwise similarity is controlled by purity, through the vector length $r$ (Eqn.~\ref{eqn:prep2}).}
\end{figure}

We define orthogonal states $\{\ket{\bm{0}},\ket{\bm{1}}\}$ that span the qubit internal space. First we consider three photons in pure states:
\begin{equation}\label{eqn:prep1}
\begin{aligned}
\ket{a}&=\cos\left(\theta/2\right)\ket{\bm{0}}+\sin\left(\theta/2\right)\ket{\bm{1}},\\
\ket{b}&=\cos\left(\theta/2\right)\ket{\bm{0}}+e^{i\frac{2\pi}{3}}\sin\left(\theta/2\right)\ket{\bm{1}},\\
\ket{c}&=\cos\left(\theta/2\right)\ket{\bm{0}}+e^{i\frac{4\pi}{3}}\sin\left(\theta/2\right)\ket{\bm{1}},
\end{aligned}
\end{equation}
with $0\leq\theta\leq\pi/2$. These vectors lie at the top of the Bloch sphere when $\theta=0$ and are equally spaced in the equator when $\theta=\pi/2$. Like the petals of a budding flower, for other values of $\theta$ they point between the top of the Bloch sphere and the equator, and remain equally spaced in azimuthal angle, as shown in Fig.~\ref{fig:fig2}b. For these pure states, $\rho_{j}=\ketbra{j}{j}, j=a,b,c$, and the three pairwise traces are equal and given by $\TR{\rho_{j}\rho_{k}}=(5+3\cos2\theta)/8$; they vary between 0.25 and 1. The volume $V_{abc}=-3\sqrt{3}/2\times\cos\theta\sin^{2}\theta$ and its magnitude varies between 0 and 1, as shown in Fig.~\ref{fig:fig2}c.

The second preparation we consider uses identical mixed states to vary the pairwise similarity solely by purity. Each photon has the same internal state:
\begin{equation}\label{eqn:prep2}
\rho_{p}=p\ketbra{\bm{0}}{\bm{0}}+(1-p)\ketbra{\bm{1}}{\bm{1}}.
\end{equation}
$p$ is the preparation probability, the length of the Bloch vector is $r=\vert 2p-1\vert$, and the state purity is $\mathcal{P}=~\frac{1}{2}\left(1+r^{2}\right)$ (see Fig.~\ref{fig:fig2}d). Here the pairwise traces are again equal and now vary between 0.5 and 1, overlapping with the range possible for the pure state configuration. However, crucially $V_{abc}$ is here always zero.

\section{Multiphoton interference statistics}
To investigate how the various distinguishing parameters manifest in interference, we consider interfering these preparations of three photons at a balanced three-port interferometer. This `tritter' is described by a unitary matrix with elements $U_{j,k}=\exp\left(jk\frac{2\pi i}{3}\right)/\sqrt{3}$. When photons with qubit internal states  labelled $a,b,c$ enter inputs $1-3$ respectively, the probabilities of various output patterns are (Appendix~\ref{appx:scatteringProbabilities}):
\begin{equation}\label{eqn:mixedDistProbs}
\begin{aligned}
P_{111}&=\frac{1}{18}\left(3+\vdd{a}{b}+\vdd{a}{c}+\vdd{b}{c}\right),\\
P_{(120)}&=\frac{1}{36}\left(3-\vdd{a}{b}-\vdd{a}{c}-\vdd{b}{c}-\sqrt{3}V_{abc}\right),\\
P_{(210)}&=\frac{1}{36}\left(3-\vdd{a}{b}-\vdd{a}{c}-\vdd{b}{c}+\sqrt{3}V_{abc}\right),\\
P_{(300)}&=\frac{2}{3}P_{111}.
\end{aligned}
\end{equation}
The subscripts indicate the numbers of photons in the individual output modes. Brackets around these output configurations denote those related by cyclic permutation of occupation numbers, so $(210)=\{210,102,021\}$. Dot products of Bloch vectors derive from the interference of paths related by pairwise exchange of photons and dependence on $V_{abc}$ comes from interfering paths related by full permutation of photons. The high symmetry of the tritter is the reason why all fully bunched probabilities are the same, and why partially bunched probabilities look similar.
The probabilities of coincidences, $P_{111}$, and of bunchings, $P_{(300)}$, depend linearly on dot products of Bloch vectors and are not sensitive $V_{abc}$. This observation and its extension to larger systems are discussed in Appendix~\ref{appx:Vabc_insensitivity}. Critically, here, the partially bunched probabilities depend on $V_{abc}$. 

\section{Experiment and results}
\begin{figure}[t]
	\centering
	\includegraphics[width=0.47\textwidth]{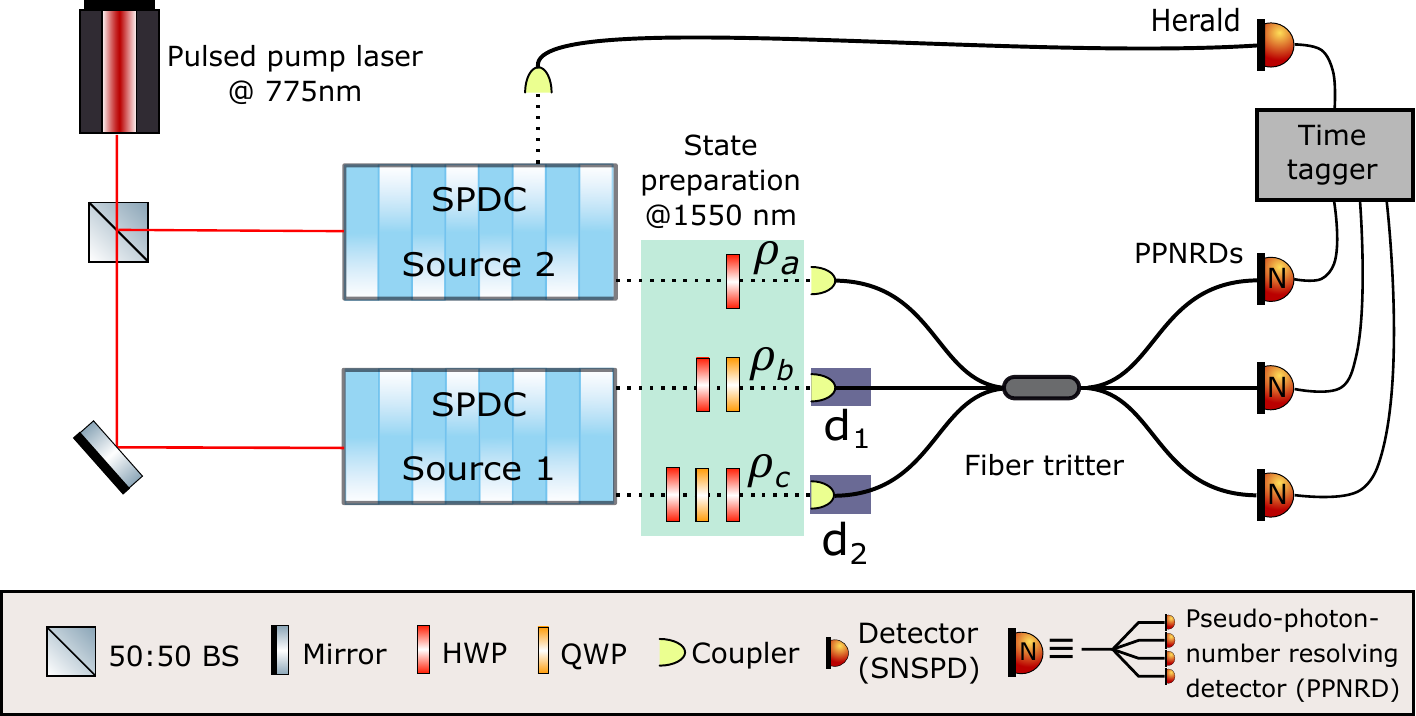}
	\caption{\label{fig:fig3} A laser pumps two SPDC sources to generate indistinguishable photons. Both photons emitted from Source 1 enter the tritter and Source 2 is operated in a heralded configuration to supply a third photon. The photons' polarisations are prepared using sets of half-wave plates (HWPs) and quarter-wave plates (QWPs) and arrival times are matched using delay stages $d_{1}$ and $d_{2}$. Each tritter output is connected to a four-port splitter for pseudo-photon-number resolution. At the measurement stage, all outputs are connected to superconducting nanowire single-photon detectors (SNSPDs) with $>90\%$ efficiencies and then a time tagger is used to count the different photon statistics.}
\end{figure}

We generate photons using spontaneous parameteric down-conversion (SPDC) and encode the qubit states in polarisation, choosing $\ket{\mathbf{0}}\coloneqq\ket{H}$ and $\ket{\mathbf{1}}\coloneqq\ket{V}$ for horizontal and vertical polarisation. Further details of the experimental setup are given in Fig.~\ref{fig:fig3} and Appendix~\ref{appx:setup}.

\begin{figure*}[ht]
	\centering
	\includegraphics[width=0.95\textwidth]{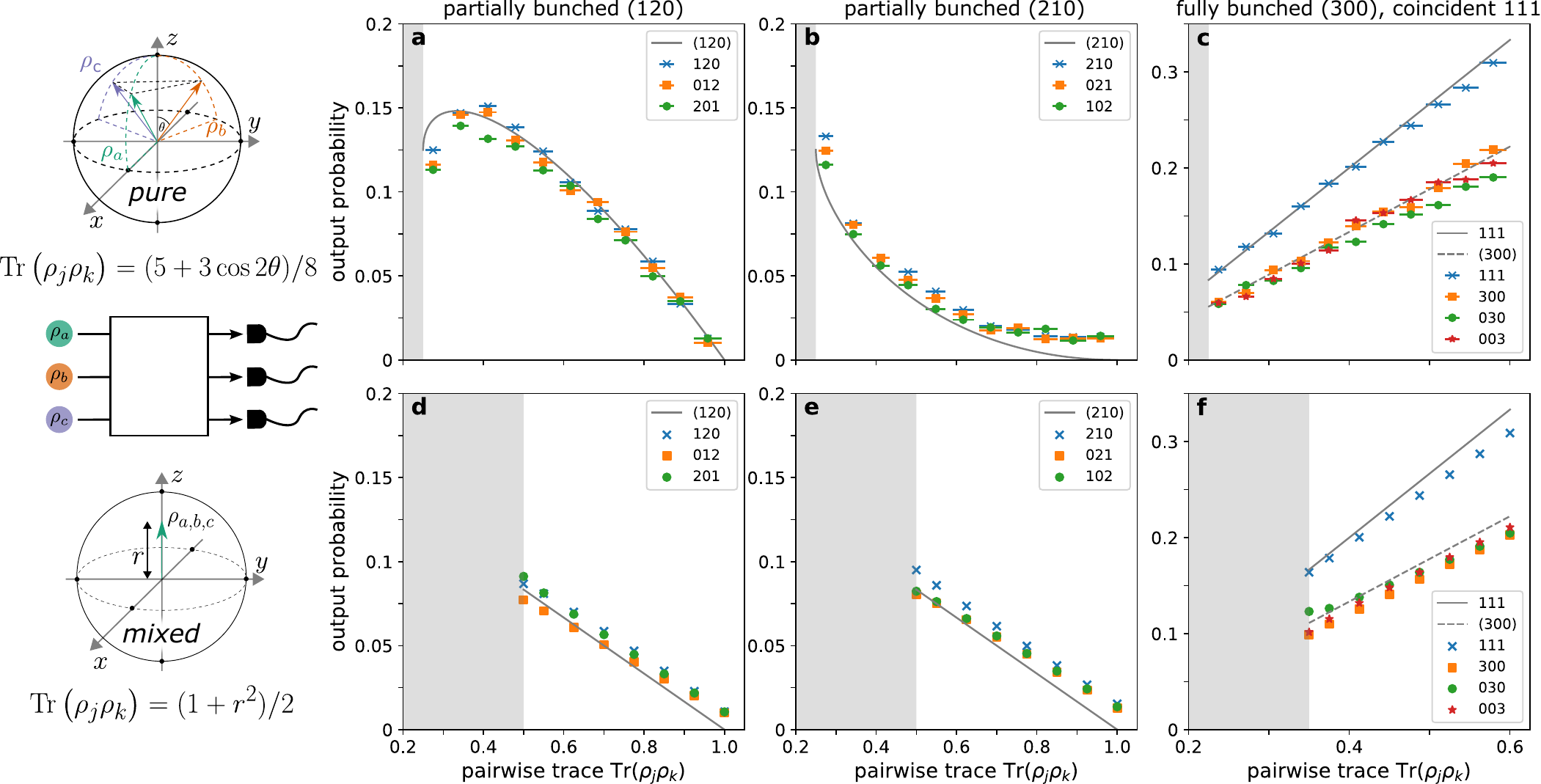}
	\caption{\label{fig:fig4} Scattering probabilities for pure (top row, \textbf{a}-\textbf{c}) and mixed (bottom row, \textbf{d}-\textbf{f}) preparations of three photons interfering in a tritter. Grey regions indicate values of the pairwise trace that cannot be accessed for each preparation, and grey curves are ideal theory. For the top row, the pairwise traces of the interfering photons vary between 1 and 0.25 as the angle $\theta$ varies from 0 to $\pi/2$. From Eqn.~\ref{eqn:mixedDistProbs}, partially bunched probabilities \textbf{a} and \textbf{b} depend differently on $V_{abc}$, but in \textbf{c} the fully bunched and coincidence probabilities depend only on pairwise traces. For the bottom row, the pairwise traces are determined by the state purity that varies between 0.5 and 1 as the Bloch vector length $r$ varies from 0 to 1. Here $V_{abc}=0$ and so the partially bunched statistics in \textbf{d} and \textbf{e} vary linearly with the pairwise trace, like the other probabilities in \textbf{f}. Horizontal error bars in the top plots are determined experimentally from mean HOM visibilities and all vertical error bars are comparable to marker size.}
\end{figure*}

We first prepare three photons in the pure states of Eqn.~\ref{eqn:prep1} (shown in Fig.~\ref{fig:fig2}b). The angle $\theta$ is varied from 0 to $\pi/2$ using waveplates. This monotonically decreases the pairwise trace $\mathrm{Tr}(\rho_{j}\rho_{k})$ from 1 to 0.25. $\theta$ also changes the volume $\vert V_{abc}\vert$ as shown in Fig.~\ref{fig:fig2}c. We then perform two sets of measurements: HOM dip visibilities between pairs of photons to infer the experimental pairwise traces, and three-photon counts at the tritter outputs to estimate the three-photon scattering probabilities. Results are shown in Fig.~\ref{fig:fig4}a-c, along with ideal theory curves. The partially bunched probabilities exhibit nonlinear variation as the pairwise traces change due to their dependence on $V_{abc}$, whereas the coincidence and fully bunched statistics vary linearly (see probabilities in Eqn.~\ref{eqn:mixedDistProbs}).

Next, we simulate the preparation of three photons in the mixed states of Eqn.~\ref{eqn:prep2} (shown in Fig.~\ref{fig:fig2}d). We achieve this by measuring three-photon counts for all eight input combinations where each photon has $H$ or $V$ polarisation. Summing these with appropriate weightings depending on preparation probability $p$ simulates the scattering probabilities for photons in identical mixed states. We select a set of pairwise traces between 0.5 and 1 for which we determine the three-photon scattering probabilities, and these are shown in Fig.~\ref{fig:fig4}d-f with ideal theory curves. The behaviour of the partially bunched probabilities now contrasts sharply with that for the pure state preparation by following a linear relation with pairwise traces because the distinguishing volume $V_{abc}=0$.

In both cases, differences between experimental results and ideal theory are due to imperfect state preparation, residual spectral distinguishability and mixedness, and higher-order photon emissions from the SPDC sources. Also, we use four-port splitters for pseudo photon number resolution and imbalances in their splitting ratio will affect the probabilities determined from photon counts.

The key takeaway from Fig.~\ref{fig:fig4} is that the two very different preparations of photons can give rise to the same HOM dip visibilities but result in very different three-photon probabilities. Both preparations exhibit pairwise traces of between 0.5 and 1  -- as indicated by overlapping regions of the shared x-axis -- and so HOM experiments cannot discriminate between them, as in Fig.~\ref{fig:fig1} and shown in additional data in Appendix~\ref{appx:HOMdata}. As expected from the expressions in Eqn.~\ref{eqn:mixedDistProbs}, the coincident and fully bunched probabilities only depend on dot products of Bloch vectors and so also cannot distinguish the state preparations, as evident in Fig.~\ref{fig:fig4}c,f.

However, Eqn.~\ref{eqn:mixedDistProbs} shows that partially bunched probabilities depend on $V_{abc}$. The mixed preparation has $V_{abc}=0$ but the pure state preparation leads to a non-zero -- and here negative -- $V_{abc}$ that can be inferred from measured probabilities. Thus for pairwise traces $0.5\leq\mathrm{Tr}(\rho_{j}\rho_{k})<1$, these measurements can tell the difference between the preparations in a way impossible using just lower-order interference. Besides discriminating between these special state preparations, measurement of HOM visibilities and $V_{abc}$ can also be used to identify mixedness of the interfering qubit internal states, as shown in Appendix~\ref{appx:identifyMixedness}.

\section{Discussion}
In this work we have presented an experiment that reveals the different effects of distinguishability and mixedness in multiphoton interference beyond two photons. The scalar triple product of Bloch vectors embodies this for qubit internal states. We can briefly comment on how the investigations here extend to more photons and higher dimensions. Adding a fourth photon with a qubit internal state $\rho_{d}$ introduces a dependence on $\TR{\rho_{a}\rho_{b}\rho_{c}\rho_{d}}$. However, as shown in Appendix~\ref{appx:fourQubits}, this depends only on dot products and scalar triple products of Bloch vectors, and so this four-photon interference is fully described by two- and three-photon parameters. This holds for the interference of any number of independent photons with qubit internal states.

The choice of qubit internal states here imposes a restriction on the five parameters governing three-photon interference (see Appendix~\ref{appx:identifyMixedness}). Turning to higher dimensions, qutrits are sufficient to freely probe all parameters. The triple trace then contains extra terms~\cite{Arvind1997,Hartley2004,Caves2000} such that, unlike for qubits, HOM visibilities do not fully determine the real part of the triple trace, as shown in Appendix~\ref{appx:qutrits}.

Our result highlights the importance of going beyond HOM visibilities when characterising photon indistinguishability. This is particularly relevant in the context of photonic quantum technologies, where much effort is dedicated to engineering sources of pure, indistinguishable photons. Common approaches include spontaneous processes that rely on material non-linearities, such as spontaneous parametric down-conversion and four-wave mixing, and in principle deterministic quantum emitters, such as quantum dots and vacancy centres in diamond. The choice of photon source will determine the dominant photon errors, necessitating careful characterisation to determine the impact on quantum operation fidelity.

Beyond tests of quantum computational complexity using photons~\cite{Zhong2020,Zhong2021}, optical approaches to universal measurement-based quantum computation rely on the generation of small entangled states that can be combined to build up a cluster resource state~\cite{Raussendorf2003,Gimeno-Segovia2015}. The effect of photon distinguishability on fault tolerant schemes has been investigated~\cite{Rohde2006}, but a more general treatment also including effects of photon impurity -- and routes to protect against such errors -- will become crucial as the scale of optical quantum technologies continues to grow. As well as photons, our work also applies to systems of other interfering particles where interactions with the environment and which-path information can degrade indistinguishability.

\bibliographystyle{apsrev4-1}
\bibliography{refs.bib}

\begin{thebibliography}{33}%
\makeatletter
\providecommand \@ifxundefined [1]{%
 \@ifx{#1\undefined}
}%
\providecommand \@ifnum [1]{%
 \ifnum #1\expandafter \@firstoftwo
 \else \expandafter \@secondoftwo
 \fi
}%
\providecommand \@ifx [1]{%
 \ifx #1\expandafter \@firstoftwo
 \else \expandafter \@secondoftwo
 \fi
}%
\providecommand \natexlab [1]{#1}%
\providecommand \enquote  [1]{``#1''}%
\providecommand \bibnamefont  [1]{#1}%
\providecommand \bibfnamefont [1]{#1}%
\providecommand \citenamefont [1]{#1}%
\providecommand \href@noop [0]{\@secondoftwo}%
\providecommand \href [0]{\begingroup \@sanitize@url \@href}%
\providecommand \@href[1]{\@@startlink{#1}\@@href}%
\providecommand \@@href[1]{\endgroup#1\@@endlink}%
\providecommand \@sanitize@url [0]{\catcode `\\12\catcode `\$12\catcode
  `\&12\catcode `\#12\catcode `\^12\catcode `\_12\catcode `\%12\relax}%
\providecommand \@@startlink[1]{}%
\providecommand \@@endlink[0]{}%
\providecommand \url  [0]{\begingroup\@sanitize@url \@url }%
\providecommand \@url [1]{\endgroup\@href {#1}{\urlprefix }}%
\providecommand \urlprefix  [0]{URL }%
\providecommand \Eprint [0]{\href }%
\providecommand \doibase [0]{http://dx.doi.org/}%
\providecommand \selectlanguage [0]{\@gobble}%
\providecommand \bibinfo  [0]{\@secondoftwo}%
\providecommand \bibfield  [0]{\@secondoftwo}%
\providecommand \translation [1]{[#1]}%
\providecommand \BibitemOpen [0]{}%
\providecommand \bibitemStop [0]{}%
\providecommand \bibitemNoStop [0]{.\EOS\space}%
\providecommand \EOS [0]{\spacefactor3000\relax}%
\providecommand \BibitemShut  [1]{\csname bibitem#1\endcsname}%
\let\auto@bib@innerbib\@empty
\bibitem [{\citenamefont {O'Brien}\ \emph {et~al.}(2009)\citenamefont
  {O'Brien}, \citenamefont {Furusawa},\ and\ \citenamefont
  {Vu{\v{c}}kovi{\'c}}}]{OBrien2009}%
  \BibitemOpen
  \bibfield  {author} {\bibinfo {author} {\bibfnamefont {J.~L.}\ \bibnamefont
  {O'Brien}}, \bibinfo {author} {\bibfnamefont {A.}~\bibnamefont {Furusawa}}, \
  and\ \bibinfo {author} {\bibfnamefont {J.}~\bibnamefont
  {Vu{\v{c}}kovi{\'c}}},\ }\href {https://doi.org/10.1038/nphoton.2009.229}
  {\bibfield  {journal} {\bibinfo  {journal} {Nature Photonics}\ }\textbf
  {\bibinfo {volume} {3}},\ \bibinfo {pages} {687} (\bibinfo {year}
  {2009})}\BibitemShut {NoStop}%
\bibitem [{\citenamefont {Kok}\ \emph {et~al.}(2007)\citenamefont {Kok},
  \citenamefont {Munro}, \citenamefont {Nemoto}, \citenamefont {Ralph},
  \citenamefont {Dowling},\ and\ \citenamefont {Milburn}}]{Kok2007}%
  \BibitemOpen
  \bibfield  {author} {\bibinfo {author} {\bibfnamefont {P.}~\bibnamefont
  {Kok}}, \bibinfo {author} {\bibfnamefont {W.~J.}\ \bibnamefont {Munro}},
  \bibinfo {author} {\bibfnamefont {K.}~\bibnamefont {Nemoto}}, \bibinfo
  {author} {\bibfnamefont {T.~C.}\ \bibnamefont {Ralph}}, \bibinfo {author}
  {\bibfnamefont {J.~P.}\ \bibnamefont {Dowling}}, \ and\ \bibinfo {author}
  {\bibfnamefont {G.~J.}\ \bibnamefont {Milburn}},\ }\href {https://doi.org/10.1103/RevModPhys.79.135} {\bibfield  {journal} {\bibinfo  {journal} {Rev.
  Mod. Phys.}\ }\textbf {\bibinfo {volume} {79}},\ \bibinfo {pages} {135}
  (\bibinfo {year} {2007})}\BibitemShut {NoStop}%
\bibitem [{\citenamefont {Zhong}\ \emph {et~al.}(2020)\citenamefont {Zhong},
  \citenamefont {Wang}, \citenamefont {Deng}, \citenamefont {Chen},
  \citenamefont {Peng}, \citenamefont {Luo}, \citenamefont {Qin}, \citenamefont
  {Wu}, \citenamefont {Ding}, \citenamefont {Hu}, \citenamefont {Hu},
  \citenamefont {Yang}, \citenamefont {Zhang}, \citenamefont {Li},
  \citenamefont {Li}, \citenamefont {Jiang}, \citenamefont {Gan}, \citenamefont
  {Yang}, \citenamefont {You}, \citenamefont {Wang}, \citenamefont {Li},
  \citenamefont {Liu}, \citenamefont {Lu},\ and\ \citenamefont
  {Pan}}]{Zhong2020}%
  \BibitemOpen
  \bibfield  {author} {\bibinfo {author} {\bibfnamefont {H.-S.}\ \bibnamefont
  {Zhong}}, \bibinfo {author} {\bibfnamefont {H.}~\bibnamefont {Wang}},
  \bibinfo {author} {\bibfnamefont {Y.-H.}\ \bibnamefont {Deng}}, \bibinfo
  {author} {\bibfnamefont {M.-C.}\ \bibnamefont {Chen}}, \bibinfo {author}
  {\bibfnamefont {L.-C.}\ \bibnamefont {Peng}}, \bibinfo {author}
  {\bibfnamefont {Y.-H.}\ \bibnamefont {Luo}}, \bibinfo {author} {\bibfnamefont
  {J.}~\bibnamefont {Qin}}, \bibinfo {author} {\bibfnamefont {D.}~\bibnamefont
  {Wu}}, \bibinfo {author} {\bibfnamefont {X.}~\bibnamefont {Ding}}, \bibinfo
  {author} {\bibfnamefont {Y.}~\bibnamefont {Hu}}, \bibinfo {author}
  {\bibfnamefont {P.}~\bibnamefont {Hu}}, \bibinfo {author} {\bibfnamefont
  {X.-Y.}\ \bibnamefont {Yang}}, \bibinfo {author} {\bibfnamefont {W.-J.}\
  \bibnamefont {Zhang}}, \bibinfo {author} {\bibfnamefont {H.}~\bibnamefont
  {Li}}, \bibinfo {author} {\bibfnamefont {Y.}~\bibnamefont {Li}}, \bibinfo
  {author} {\bibfnamefont {X.}~\bibnamefont {Jiang}}, \bibinfo {author}
  {\bibfnamefont {L.}~\bibnamefont {Gan}}, \bibinfo {author} {\bibfnamefont
  {G.}~\bibnamefont {Yang}}, \bibinfo {author} {\bibfnamefont {L.}~\bibnamefont
  {You}}, \bibinfo {author} {\bibfnamefont {Z.}~\bibnamefont {Wang}}, \bibinfo
  {author} {\bibfnamefont {L.}~\bibnamefont {Li}}, \bibinfo {author}
  {\bibfnamefont {N.-L.}\ \bibnamefont {Liu}}, \bibinfo {author} {\bibfnamefont
  {C.-Y.}\ \bibnamefont {Lu}}, \ and\ \bibinfo {author} {\bibfnamefont {J.-W.}\
  \bibnamefont {Pan}},\ }\href {\doibase 10.1126/science.abe8770} {\bibfield
  {journal} {\bibinfo  {journal} {Science}\ }\textbf {\bibinfo {volume}
  {370}},\ \bibinfo {pages} {1460} (\bibinfo {year} {2020})}\BibitemShut
  {NoStop}%
\bibitem [{\citenamefont {Polino}\ \emph {et~al.}(2020)\citenamefont {Polino},
  \citenamefont {Valeri}, \citenamefont {Spagnolo},\ and\ \citenamefont
  {Sciarrino}}]{Polino2020}%
  \BibitemOpen
  \bibfield  {author} {\bibinfo {author} {\bibfnamefont {E.}~\bibnamefont
  {Polino}}, \bibinfo {author} {\bibfnamefont {M.}~\bibnamefont {Valeri}},
  \bibinfo {author} {\bibfnamefont {N.}~\bibnamefont {Spagnolo}}, \ and\
  \bibinfo {author} {\bibfnamefont {F.}~\bibnamefont {Sciarrino}},\ }\href
  {\doibase 10.1116/5.0007577} {\bibfield  {journal} {\bibinfo  {journal} {AVS
  Quantum Science}\ }\textbf {\bibinfo {volume} {2}},\ \bibinfo {pages}
  {024703} (\bibinfo {year} {2020})}\BibitemShut {NoStop}%
\bibitem [{\citenamefont {Sangouard}\ \emph {et~al.}(2011)\citenamefont
  {Sangouard}, \citenamefont {Simon}, \citenamefont {de~Riedmatten},\ and\
  \citenamefont {Gisin}}]{Sangouard2011}%
  \BibitemOpen
  \bibfield  {author} {\bibinfo {author} {\bibfnamefont {N.}~\bibnamefont
  {Sangouard}}, \bibinfo {author} {\bibfnamefont {C.}~\bibnamefont {Simon}},
  \bibinfo {author} {\bibfnamefont {H.}~\bibnamefont {de~Riedmatten}}, \ and\
  \bibinfo {author} {\bibfnamefont {N.}~\bibnamefont {Gisin}},\ }\href
  {\doibase 10.1103/RevModPhys.83.33} {\bibfield  {journal} {\bibinfo
  {journal} {Rev. Mod. Phys.}\ }\textbf {\bibinfo {volume} {83}},\ \bibinfo
  {pages} {33} (\bibinfo {year} {2011})}\BibitemShut {NoStop}%
\bibitem [{\citenamefont {Kimble}(2008)}]{Kimble2008}%
  \BibitemOpen
  \bibfield  {author} {\bibinfo {author} {\bibfnamefont {H.~J.}\ \bibnamefont
  {Kimble}},\ }\href {\doibase 10.1038/nature07127} {\bibfield  {journal}
  {\bibinfo  {journal} {Nature}\ }\textbf {\bibinfo {volume} {453}},\ \bibinfo
  {pages} {1023} (\bibinfo {year} {2008})}\BibitemShut {NoStop}%
\bibitem [{\citenamefont {Pompili}\ \emph {et~al.}(2021)\citenamefont
  {Pompili}, \citenamefont {Hermans}, \citenamefont {Baier}, \citenamefont
  {Beukers}, \citenamefont {Humphreys}, \citenamefont {Schouten}, \citenamefont
  {Vermeulen}, \citenamefont {Tiggelman}, \citenamefont {dos Santos~Martins},
  \citenamefont {Dirkse}, \citenamefont {Wehner},\ and\ \citenamefont
  {Hanson}}]{Pompili2021}%
  \BibitemOpen
  \bibfield  {author} {\bibinfo {author} {\bibfnamefont {M.}~\bibnamefont
  {Pompili}}, \bibinfo {author} {\bibfnamefont {S.~L.~N.}\ \bibnamefont
  {Hermans}}, \bibinfo {author} {\bibfnamefont {S.}~\bibnamefont {Baier}},
  \bibinfo {author} {\bibfnamefont {H.~K.~C.}\ \bibnamefont {Beukers}},
  \bibinfo {author} {\bibfnamefont {P.~C.}\ \bibnamefont {Humphreys}}, \bibinfo
  {author} {\bibfnamefont {R.~N.}\ \bibnamefont {Schouten}}, \bibinfo {author}
  {\bibfnamefont {R.~F.~L.}\ \bibnamefont {Vermeulen}}, \bibinfo {author}
  {\bibfnamefont {M.~J.}\ \bibnamefont {Tiggelman}}, \bibinfo {author}
  {\bibfnamefont {L.}~\bibnamefont {dos Santos~Martins}}, \bibinfo {author}
  {\bibfnamefont {B.}~\bibnamefont {Dirkse}}, \bibinfo {author} {\bibfnamefont
  {S.}~\bibnamefont {Wehner}}, \ and\ \bibinfo {author} {\bibfnamefont
  {R.}~\bibnamefont {Hanson}},\ }\href {\doibase 10.1126/science.abg1919}
  {\bibfield  {journal} {\bibinfo  {journal} {Science}\ }\textbf {\bibinfo
  {volume} {372}},\ \bibinfo {pages} {259} (\bibinfo {year}
  {2021})}\BibitemShut {NoStop}%
\bibitem [{\citenamefont {Englert}(1996)}]{Englert1996}%
  \BibitemOpen
  \bibfield  {author} {\bibinfo {author} {\bibfnamefont {B.-G.}\ \bibnamefont
  {Englert}},\ }\href {\doibase 10.1103/PhysRevLett.77.2154} {\bibfield
  {journal} {\bibinfo  {journal} {Phys. Rev. Lett.}\ }\textbf {\bibinfo
  {volume} {77}},\ \bibinfo {pages} {2154} (\bibinfo {year}
  {1996})}\BibitemShut {NoStop}%
\bibitem [{\citenamefont {Dittel}\ \emph {et~al.}(2021)\citenamefont {Dittel},
  \citenamefont {Dufour}, \citenamefont {Weihs},\ and\ \citenamefont
  {Buchleitner}}]{Dittel2021}%
  \BibitemOpen
  \bibfield  {author} {\bibinfo {author} {\bibfnamefont {C.}~\bibnamefont
  {Dittel}}, \bibinfo {author} {\bibfnamefont {G.}~\bibnamefont {Dufour}},
  \bibinfo {author} {\bibfnamefont {G.}~\bibnamefont {Weihs}}, \ and\ \bibinfo
  {author} {\bibfnamefont {A.}~\bibnamefont {Buchleitner}},\ }\href {\doibase 10.1103/PhysRevX.11.031041} {\bibfield  {journal} {\bibinfo  {journal} {Phys.
  Rev. X}\ }\textbf {\bibinfo {volume} {11}},\ \bibinfo {pages} {031041}
  (\bibinfo {year} {2021})}\BibitemShut {NoStop}%
\bibitem [{\citenamefont {Tichy}(2015)}]{Tichy2015}%
  \BibitemOpen
  \bibfield  {author} {\bibinfo {author} {\bibfnamefont {M.~C.}\ \bibnamefont
  {Tichy}},\ }\href {\doibase 10.1103/PhysRevA.91.022316} {\bibfield  {journal}
  {\bibinfo  {journal} {Phys. Rev. A}\ }\textbf {\bibinfo {volume} {91}},\
  \bibinfo {pages} {022316} (\bibinfo {year} {2015})}\BibitemShut {NoStop}%
\bibitem [{\citenamefont {Tillmann}\ \emph {et~al.}(2015)\citenamefont
  {Tillmann}, \citenamefont {Tan}, \citenamefont {Stoeckl}, \citenamefont
  {Sanders}, \citenamefont {de~Guise}, \citenamefont {Heilmann}, \citenamefont
  {Nolte}, \citenamefont {Szameit},\ and\ \citenamefont
  {Walther}}]{Tillmann2015}%
  \BibitemOpen
  \bibfield  {author} {\bibinfo {author} {\bibfnamefont {M.}~\bibnamefont
  {Tillmann}}, \bibinfo {author} {\bibfnamefont {S.-H.}\ \bibnamefont {Tan}},
  \bibinfo {author} {\bibfnamefont {S.~E.}\ \bibnamefont {Stoeckl}}, \bibinfo
  {author} {\bibfnamefont {B.~C.}\ \bibnamefont {Sanders}}, \bibinfo {author}
  {\bibfnamefont {H.}~\bibnamefont {de~Guise}}, \bibinfo {author}
  {\bibfnamefont {R.}~\bibnamefont {Heilmann}}, \bibinfo {author}
  {\bibfnamefont {S.}~\bibnamefont {Nolte}}, \bibinfo {author} {\bibfnamefont
  {A.}~\bibnamefont {Szameit}}, \ and\ \bibinfo {author} {\bibfnamefont
  {P.}~\bibnamefont {Walther}},\ }\href {\doibase 10.1103/PhysRevX.5.041015}
  {\bibfield  {journal} {\bibinfo  {journal} {Phys. Rev. X}\ }\textbf {\bibinfo
  {volume} {5}},\ \bibinfo {pages} {041015} (\bibinfo {year}
  {2015})}\BibitemShut {NoStop}%
\bibitem [{\citenamefont {Shchesnovich}(2015)}]{Shchesnovich2015}%
  \BibitemOpen
  \bibfield  {author} {\bibinfo {author} {\bibfnamefont {V.~S.}\ \bibnamefont
  {Shchesnovich}},\ }\href {\doibase 10.1103/PhysRevA.91.013844} {\bibfield
  {journal} {\bibinfo  {journal} {Phys. Rev. A}\ }\textbf {\bibinfo {volume}
  {91}},\ \bibinfo {pages} {013844} (\bibinfo {year} {2015})}\BibitemShut
  {NoStop}%
\bibitem [{\citenamefont {Shchesnovich}\ and\ \citenamefont
  {Bezerra}(2018)}]{Shchesnovich2018}%
  \BibitemOpen
  \bibfield  {author} {\bibinfo {author} {\bibfnamefont {V.~S.}\ \bibnamefont
  {Shchesnovich}}\ and\ \bibinfo {author} {\bibfnamefont {M.~E.~O.}\
  \bibnamefont {Bezerra}},\ }\href {\doibase 10.1103/PhysRevA.98.033805}
  {\bibfield  {journal} {\bibinfo  {journal} {Phys. Rev. A}\ }\textbf {\bibinfo
  {volume} {98}},\ \bibinfo {pages} {033805} (\bibinfo {year}
  {2018})}\BibitemShut {NoStop}%
\bibitem [{\citenamefont {Renema}\ \emph {et~al.}(2018)\citenamefont {Renema},
  \citenamefont {Menssen}, \citenamefont {Clements}, \citenamefont {Triginer},
  \citenamefont {Kolthammer},\ and\ \citenamefont {Walmsley}}]{Renema2018}%
  \BibitemOpen
  \bibfield  {author} {\bibinfo {author} {\bibfnamefont {J.~J.}\ \bibnamefont
  {Renema}}, \bibinfo {author} {\bibfnamefont {A.}~\bibnamefont {Menssen}},
  \bibinfo {author} {\bibfnamefont {W.~R.}\ \bibnamefont {Clements}}, \bibinfo
  {author} {\bibfnamefont {G.}~\bibnamefont {Triginer}}, \bibinfo {author}
  {\bibfnamefont {W.~S.}\ \bibnamefont {Kolthammer}}, \ and\ \bibinfo {author}
  {\bibfnamefont {I.~A.}\ \bibnamefont {Walmsley}},\ }\href {\doibase 10.1103/PhysRevLett.120.220502} {\bibfield  {journal} {\bibinfo  {journal}
  {Phys. Rev. Lett.}\ }\textbf {\bibinfo {volume} {120}},\ \bibinfo {pages}
  {220502} (\bibinfo {year} {2018})}\BibitemShut {NoStop}%
\bibitem [{\citenamefont {Menssen}\ \emph {et~al.}(2017)\citenamefont
  {Menssen}, \citenamefont {Jones}, \citenamefont {Metcalf}, \citenamefont
  {Tichy}, \citenamefont {Barz}, \citenamefont {Kolthammer},\ and\
  \citenamefont {Walmsley}}]{Menssen2017}%
  \BibitemOpen
  \bibfield  {author} {\bibinfo {author} {\bibfnamefont {A.~J.}\ \bibnamefont
  {Menssen}}, \bibinfo {author} {\bibfnamefont {A.~E.}\ \bibnamefont {Jones}},
  \bibinfo {author} {\bibfnamefont {B.~J.}\ \bibnamefont {Metcalf}}, \bibinfo
  {author} {\bibfnamefont {M.~C.}\ \bibnamefont {Tichy}}, \bibinfo {author}
  {\bibfnamefont {S.}~\bibnamefont {Barz}}, \bibinfo {author} {\bibfnamefont
  {W.~S.}\ \bibnamefont {Kolthammer}}, \ and\ \bibinfo {author} {\bibfnamefont
  {I.~A.}\ \bibnamefont {Walmsley}},\ }\href {\doibase 10.1103/PhysRevLett.118.153603} {\bibfield  {journal} {\bibinfo  {journal}
  {Phys. Rev. Lett.}\ }\textbf {\bibinfo {volume} {118}},\ \bibinfo {pages}
  {153603} (\bibinfo {year} {2017})}\BibitemShut {NoStop}%
\bibitem [{\citenamefont {Jones}\ \emph {et~al.}(2020)\citenamefont {Jones},
  \citenamefont {Menssen}, \citenamefont {Chrzanowski}, \citenamefont
  {Wolterink}, \citenamefont {Shchesnovich},\ and\ \citenamefont
  {Walmsley}}]{Jones2020}%
  \BibitemOpen
  \bibfield  {author} {\bibinfo {author} {\bibfnamefont {A.~E.}\ \bibnamefont
  {Jones}}, \bibinfo {author} {\bibfnamefont {A.~J.}\ \bibnamefont {Menssen}},
  \bibinfo {author} {\bibfnamefont {H.~M.}\ \bibnamefont {Chrzanowski}},
  \bibinfo {author} {\bibfnamefont {T.~A.~W.}\ \bibnamefont {Wolterink}},
  \bibinfo {author} {\bibfnamefont {V.~S.}\ \bibnamefont {Shchesnovich}}, \
  and\ \bibinfo {author} {\bibfnamefont {I.~A.}\ \bibnamefont {Walmsley}},\
  }\href {\doibase 10.1103/PhysRevLett.125.123603} {\bibfield  {journal}
  {\bibinfo  {journal} {Phys. Rev. Lett.}\ }\textbf {\bibinfo {volume} {125}},\
  \bibinfo {pages} {123603} (\bibinfo {year} {2020})}\BibitemShut {NoStop}%
\bibitem [{\citenamefont {Minke}\ \emph {et~al.}(2021)\citenamefont {Minke},
  \citenamefont {Buchleitner},\ and\ \citenamefont {Dittel}}]{Minke2021}%
  \BibitemOpen
  \bibfield  {author} {\bibinfo {author} {\bibfnamefont {A.~M.}\ \bibnamefont
  {Minke}}, \bibinfo {author} {\bibfnamefont {A.}~\bibnamefont {Buchleitner}},
  \ and\ \bibinfo {author} {\bibfnamefont {C.}~\bibnamefont {Dittel}},\ }\href
  {\doibase 10.1088/1367-2630/ac0fb1} {\bibfield  {journal} {\bibinfo
  {journal} {New Journal of Physics}\ }\textbf {\bibinfo {volume} {23}},\
  \bibinfo {pages} {073028} (\bibinfo {year} {2021})}\BibitemShut {NoStop}%
\bibitem [{\citenamefont {Brod}\ \emph {et~al.}(2019)\citenamefont {Brod},
  \citenamefont {Galv\~ao}, \citenamefont {Viggianiello}, \citenamefont
  {Flamini}, \citenamefont {Spagnolo},\ and\ \citenamefont
  {Sciarrino}}]{Brod2019}%
  \BibitemOpen
  \bibfield  {author} {\bibinfo {author} {\bibfnamefont {D.~J.}\ \bibnamefont
  {Brod}}, \bibinfo {author} {\bibfnamefont {E.~F.}\ \bibnamefont {Galv\~ao}},
  \bibinfo {author} {\bibfnamefont {N.}~\bibnamefont {Viggianiello}}, \bibinfo
  {author} {\bibfnamefont {F.}~\bibnamefont {Flamini}}, \bibinfo {author}
  {\bibfnamefont {N.}~\bibnamefont {Spagnolo}}, \ and\ \bibinfo {author}
  {\bibfnamefont {F.}~\bibnamefont {Sciarrino}},\ }\href {\doibase 10.1103/PhysRevLett.122.063602} {\bibfield  {journal} {\bibinfo  {journal}
  {Phys. Rev. Lett.}\ }\textbf {\bibinfo {volume} {122}},\ \bibinfo {pages}
  {063602} (\bibinfo {year} {2019})}\BibitemShut {NoStop}%
\bibitem [{\citenamefont {Viggianiello}\ \emph {et~al.}(2018)\citenamefont
  {Viggianiello}, \citenamefont {Flamini}, \citenamefont {Bentivegna},
  \citenamefont {Spagnolo}, \citenamefont {Crespi}, \citenamefont {Brod},
  \citenamefont {Galvão}, \citenamefont {Osellame},\ and\ \citenamefont
  {Sciarrino}}]{Viggianiello2018}%
  \BibitemOpen
  \bibfield  {author} {\bibinfo {author} {\bibfnamefont {N.}~\bibnamefont
  {Viggianiello}}, \bibinfo {author} {\bibfnamefont {F.}~\bibnamefont
  {Flamini}}, \bibinfo {author} {\bibfnamefont {M.}~\bibnamefont {Bentivegna}},
  \bibinfo {author} {\bibfnamefont {N.}~\bibnamefont {Spagnolo}}, \bibinfo
  {author} {\bibfnamefont {A.}~\bibnamefont {Crespi}}, \bibinfo {author}
  {\bibfnamefont {D.~J.}\ \bibnamefont {Brod}}, \bibinfo {author}
  {\bibfnamefont {E.~F.}\ \bibnamefont {Galvão}}, \bibinfo {author}
  {\bibfnamefont {R.}~\bibnamefont {Osellame}}, \ and\ \bibinfo {author}
  {\bibfnamefont {F.}~\bibnamefont {Sciarrino}},\ }\href {\doibase
  https://doi.org/10.1016/j.scib.2018.10.009} {\bibfield  {journal} {\bibinfo
  {journal} {Science Bulletin}\ }\textbf {\bibinfo {volume} {63}},\ \bibinfo
  {pages} {1470} (\bibinfo {year} {2018})}\BibitemShut {NoStop}%
\bibitem [{\citenamefont {Hong}\ \emph {et~al.}(1987)\citenamefont {Hong},
  \citenamefont {Ou},\ and\ \citenamefont {Mandel}}]{Hong1987}%
  \BibitemOpen
  \bibfield  {author} {\bibinfo {author} {\bibfnamefont {C.~K.}\ \bibnamefont
  {Hong}}, \bibinfo {author} {\bibfnamefont {Z.~Y.}\ \bibnamefont {Ou}}, \ and\
  \bibinfo {author} {\bibfnamefont {L.}~\bibnamefont {Mandel}},\ }\href
  {\doibase 10.1103/PhysRevLett.59.2044} {\bibfield  {journal} {\bibinfo
  {journal} {Phys. Rev. Lett.}\ }\textbf {\bibinfo {volume} {59}},\ \bibinfo
  {pages} {2044} (\bibinfo {year} {1987})}\BibitemShut {NoStop}%
\bibitem [{\citenamefont {Bouchard}\ \emph {et~al.}(2021)\citenamefont
  {Bouchard}, \citenamefont {Sit}, \citenamefont {Zhang}, \citenamefont
  {Fickler}, \citenamefont {Miatto}, \citenamefont {Yao}, \citenamefont
  {Sciarrino},\ and\ \citenamefont {Karimi}}]{Bouchard2021}%
  \BibitemOpen
  \bibfield  {author} {\bibinfo {author} {\bibfnamefont {F.}~\bibnamefont
  {Bouchard}}, \bibinfo {author} {\bibfnamefont {A.}~\bibnamefont {Sit}},
  \bibinfo {author} {\bibfnamefont {Y.}~\bibnamefont {Zhang}}, \bibinfo
  {author} {\bibfnamefont {R.}~\bibnamefont {Fickler}}, \bibinfo {author}
  {\bibfnamefont {F.~M.}\ \bibnamefont {Miatto}}, \bibinfo {author}
  {\bibfnamefont {Y.}~\bibnamefont {Yao}}, \bibinfo {author} {\bibfnamefont
  {F.}~\bibnamefont {Sciarrino}}, \ and\ \bibinfo {author} {\bibfnamefont
  {E.}~\bibnamefont {Karimi}},\ }\href {\doibase 10.1088/1361-6633/abcd7a}
  {\bibfield  {journal} {\bibinfo  {journal} {Reports on Progress in Physics}\
  }\textbf {\bibinfo {volume} {84}},\ \bibinfo {pages} {012402} (\bibinfo
  {year} {2021})}\BibitemShut {NoStop}%
\bibitem [{\citenamefont {Rohde}(2015)}]{Rohde2015}%
  \BibitemOpen
  \bibfield  {author} {\bibinfo {author} {\bibfnamefont {P.~P.}\ \bibnamefont
  {Rohde}},\ }\href {\doibase 10.1103/PhysRevA.91.012307} {\bibfield  {journal}
  {\bibinfo  {journal} {Phys. Rev. A}\ }\textbf {\bibinfo {volume} {91}},\
  \bibinfo {pages} {012307} (\bibinfo {year} {2015})}\BibitemShut {NoStop}%
\bibitem [{\citenamefont {Stanisic}\ and\ \citenamefont
  {Turner}(2018)}]{Stanisic2018}%
  \BibitemOpen
  \bibfield  {author} {\bibinfo {author} {\bibfnamefont {S.}~\bibnamefont
  {Stanisic}}\ and\ \bibinfo {author} {\bibfnamefont {P.~S.}\ \bibnamefont
  {Turner}},\ }\href {\doibase 10.1103/PhysRevA.98.043839} {\bibfield
  {journal} {\bibinfo  {journal} {Phys. Rev. A}\ }\textbf {\bibinfo {volume}
  {98}},\ \bibinfo {pages} {043839} (\bibinfo {year} {2018})}\BibitemShut
  {NoStop}%
\bibitem [{\citenamefont {Arvind}\ \emph
  {et~al.}(1997{\natexlab{a}})\citenamefont {Arvind}, \citenamefont {Mallesh},\
  and\ \citenamefont {Mukunda}}]{Arvind1997}%
  \BibitemOpen
  \bibfield  {author} {\bibinfo {author} {\bibnamefont {Arvind}}, \bibinfo
  {author} {\bibfnamefont {K.~S.}\ \bibnamefont {Mallesh}}, \ and\ \bibinfo
  {author} {\bibfnamefont {N.}~\bibnamefont {Mukunda}},\ }\href {\doibase 10.1088/0305-4470/30/7/021} {\bibfield  {journal} {\bibinfo  {journal}
  {Journal of Physics A: Mathematical and General}\ }\textbf {\bibinfo {volume}
  {30}},\ \bibinfo {pages} {2417} (\bibinfo {year}
  {1997}{\natexlab{a}})}\BibitemShut {NoStop}%
\bibitem [{\citenamefont {Hartley}\ and\ \citenamefont
  {Vedral}(2004)}]{Hartley2004}%
  \BibitemOpen
  \bibfield  {author} {\bibinfo {author} {\bibfnamefont {J.}~\bibnamefont
  {Hartley}}\ and\ \bibinfo {author} {\bibfnamefont {V.}~\bibnamefont
  {Vedral}},\ }\href {\doibase 10.1088/0305-4470/37/46/011} {\bibfield
  {journal} {\bibinfo  {journal} {Journal of Physics A: Mathematical and
  General}\ }\textbf {\bibinfo {volume} {37}},\ \bibinfo {pages} {11259}
  (\bibinfo {year} {2004})}\BibitemShut {NoStop}%
\bibitem [{\citenamefont {Caves}\ and\ \citenamefont
  {Milburn}(2000)}]{Caves2000}%
  \BibitemOpen
  \bibfield  {author} {\bibinfo {author} {\bibfnamefont {C.~M.}\ \bibnamefont
  {Caves}}\ and\ \bibinfo {author} {\bibfnamefont {G.~J.}\ \bibnamefont
  {Milburn}},\ }\href {\doibase 10.1016/s0030-4018(99)00693-8} {\bibfield
  {journal} {\bibinfo  {journal} {Optics Communications}\ }\textbf {\bibinfo
  {volume} {179}},\ \bibinfo {pages} {439–446} (\bibinfo {year}
  {2000})}\BibitemShut {NoStop}%
\bibitem [{\citenamefont {Zhong}\ \emph {et~al.}(2021)\citenamefont {Zhong},
  \citenamefont {Deng}, \citenamefont {Qin}, \citenamefont {Wang},
  \citenamefont {Chen}, \citenamefont {Peng}, \citenamefont {Luo},
  \citenamefont {Wu}, \citenamefont {Gong}, \citenamefont {Su}, \citenamefont
  {Hu}, \citenamefont {Hu}, \citenamefont {Yang}, \citenamefont {Zhang},
  \citenamefont {Li}, \citenamefont {Li}, \citenamefont {Jiang}, \citenamefont
  {Gan}, \citenamefont {Yang}, \citenamefont {You}, \citenamefont {Wang},
  \citenamefont {Li}, \citenamefont {Liu}, \citenamefont {Renema},
  \citenamefont {Lu},\ and\ \citenamefont {Pan}}]{Zhong2021}%
  \BibitemOpen
  \bibfield  {author} {\bibinfo {author} {\bibfnamefont {H.-S.}\ \bibnamefont
  {Zhong}}, \bibinfo {author} {\bibfnamefont {Y.-H.}\ \bibnamefont {Deng}},
  \bibinfo {author} {\bibfnamefont {J.}~\bibnamefont {Qin}}, \bibinfo {author}
  {\bibfnamefont {H.}~\bibnamefont {Wang}}, \bibinfo {author} {\bibfnamefont
  {M.-C.}\ \bibnamefont {Chen}}, \bibinfo {author} {\bibfnamefont {L.-C.}\
  \bibnamefont {Peng}}, \bibinfo {author} {\bibfnamefont {Y.-H.}\ \bibnamefont
  {Luo}}, \bibinfo {author} {\bibfnamefont {D.}~\bibnamefont {Wu}}, \bibinfo
  {author} {\bibfnamefont {S.-Q.}\ \bibnamefont {Gong}}, \bibinfo {author}
  {\bibfnamefont {H.}~\bibnamefont {Su}}, \bibinfo {author} {\bibfnamefont
  {Y.}~\bibnamefont {Hu}}, \bibinfo {author} {\bibfnamefont {P.}~\bibnamefont
  {Hu}}, \bibinfo {author} {\bibfnamefont {X.-Y.}\ \bibnamefont {Yang}},
  \bibinfo {author} {\bibfnamefont {W.-J.}\ \bibnamefont {Zhang}}, \bibinfo
  {author} {\bibfnamefont {H.}~\bibnamefont {Li}}, \bibinfo {author}
  {\bibfnamefont {Y.}~\bibnamefont {Li}}, \bibinfo {author} {\bibfnamefont
  {X.}~\bibnamefont {Jiang}}, \bibinfo {author} {\bibfnamefont
  {L.}~\bibnamefont {Gan}}, \bibinfo {author} {\bibfnamefont {G.}~\bibnamefont
  {Yang}}, \bibinfo {author} {\bibfnamefont {L.}~\bibnamefont {You}}, \bibinfo
  {author} {\bibfnamefont {Z.}~\bibnamefont {Wang}}, \bibinfo {author}
  {\bibfnamefont {L.}~\bibnamefont {Li}}, \bibinfo {author} {\bibfnamefont
  {N.-L.}\ \bibnamefont {Liu}}, \bibinfo {author} {\bibfnamefont {J.~J.}\
  \bibnamefont {Renema}}, \bibinfo {author} {\bibfnamefont {C.-Y.}\
  \bibnamefont {Lu}}, \ and\ \bibinfo {author} {\bibfnamefont {J.-W.}\
  \bibnamefont {Pan}},\ }\href {\doibase 10.1103/PhysRevLett.127.180502}
  {\bibfield  {journal} {\bibinfo  {journal} {Phys. Rev. Lett.}\ }\textbf
  {\bibinfo {volume} {127}},\ \bibinfo {pages} {180502} (\bibinfo {year}
  {2021})}\BibitemShut {NoStop}%
\bibitem [{\citenamefont {Raussendorf}\ \emph {et~al.}(2003)\citenamefont
  {Raussendorf}, \citenamefont {Browne},\ and\ \citenamefont
  {Briegel}}]{Raussendorf2003}%
  \BibitemOpen
  \bibfield  {author} {\bibinfo {author} {\bibfnamefont {R.}~\bibnamefont
  {Raussendorf}}, \bibinfo {author} {\bibfnamefont {D.~E.}\ \bibnamefont
  {Browne}}, \ and\ \bibinfo {author} {\bibfnamefont {H.~J.}\ \bibnamefont
  {Briegel}},\ }\href {\doibase 10.1103/PhysRevA.68.022312} {\bibfield
  {journal} {\bibinfo  {journal} {Phys. Rev. A}\ }\textbf {\bibinfo {volume}
  {68}},\ \bibinfo {pages} {022312} (\bibinfo {year} {2003})}\BibitemShut
  {NoStop}%
\bibitem [{\citenamefont {Gimeno-Segovia}\ \emph {et~al.}(2015)\citenamefont
  {Gimeno-Segovia}, \citenamefont {Shadbolt}, \citenamefont {Browne},\ and\
  \citenamefont {Rudolph}}]{Gimeno-Segovia2015}%
  \BibitemOpen
  \bibfield  {author} {\bibinfo {author} {\bibfnamefont {M.}~\bibnamefont
  {Gimeno-Segovia}}, \bibinfo {author} {\bibfnamefont {P.}~\bibnamefont
  {Shadbolt}}, \bibinfo {author} {\bibfnamefont {D.~E.}\ \bibnamefont
  {Browne}}, \ and\ \bibinfo {author} {\bibfnamefont {T.}~\bibnamefont
  {Rudolph}},\ }\href {\doibase 10.1103/PhysRevLett.115.020502} {\bibfield
  {journal} {\bibinfo  {journal} {Phys. Rev. Lett.}\ }\textbf {\bibinfo
  {volume} {115}},\ \bibinfo {pages} {020502} (\bibinfo {year}
  {2015})}\BibitemShut {NoStop}%
\bibitem [{\citenamefont {Rohde}\ and\ \citenamefont
  {Ralph}(2006)}]{Rohde2006}%
  \BibitemOpen
  \bibfield  {author} {\bibinfo {author} {\bibfnamefont {P.~P.}\ \bibnamefont
  {Rohde}}\ and\ \bibinfo {author} {\bibfnamefont {T.~C.}\ \bibnamefont
  {Ralph}},\ }\href {\doibase 10.1103/PhysRevA.73.062312} {\bibfield  {journal}
  {\bibinfo  {journal} {Phys. Rev. A}\ }\textbf {\bibinfo {volume} {73}},\
  \bibinfo {pages} {062312} (\bibinfo {year} {2006})}\BibitemShut {NoStop}%
\bibitem [{\citenamefont {Tilma}\ and\ \citenamefont
  {Sudarshan}(2002)}]{Tilma_2002}%
  \BibitemOpen
  \bibfield  {author} {\bibinfo {author} {\bibfnamefont {T.}~\bibnamefont
  {Tilma}}\ and\ \bibinfo {author} {\bibfnamefont {E.~C.~G.}\ \bibnamefont
  {Sudarshan}},\ }\href {\doibase 10.1088/0305-4470/35/48/316} {\bibfield
  {journal} {\bibinfo  {journal} {Journal of Physics A: Mathematical and
  General}\ }\textbf {\bibinfo {volume} {35}},\ \bibinfo {pages} {10467}
  (\bibinfo {year} {2002})}\BibitemShut {NoStop}%
\bibitem [{\citenamefont {Casey}(1889)}]{Casey1889}%
  \BibitemOpen
  \bibfield  {author} {\bibinfo {author} {\bibfnamefont {J.}~\bibnamefont
  {Casey}},\ }\href@noop {} {\emph {\bibinfo {title} {{A Treatise on Spherical
  Trigonometry}}}}\ (\bibinfo  {publisher} {Hodges, Figgis {\&} Co.},\ \bibinfo
  {address} {Dublin},\ \bibinfo {year} {1889})\BibitemShut {NoStop}%
\bibitem [{\citenamefont {Arvind}\ \emph
  {et~al.}(1997{\natexlab{b}})\citenamefont {Arvind}, \citenamefont {Mallesh},\
  and\ \citenamefont {Mukunda}}]{Arvind_1997}%
  \BibitemOpen
  \bibfield  {author} {\bibinfo {author} {\bibnamefont {Arvind}}, \bibinfo
  {author} {\bibfnamefont {K.~S.}\ \bibnamefont {Mallesh}}, \ and\ \bibinfo
  {author} {\bibfnamefont {N.}~\bibnamefont {Mukunda}},\ }\href {\doibase
  10.1088/0305-4470/30/7/021} {\bibfield  {journal} {\bibinfo  {journal}
  {Journal of Physics A: Mathematical and General}\ }\textbf {\bibinfo {volume}
  {30}},\ \bibinfo {pages} {2417} (\bibinfo {year}
  {1997}{\natexlab{b}})}\BibitemShut {NoStop}%
\end{thebibliography}%

\begin{footnotesize}
\noindent
\\
\noindent\textbf{Acknowledgements}
\noindent We thank M. Tichy and H. Chrzanowski for useful discussions. We acknowledge support from the Carl Zeiss Foundation, the Centre for Integrated Quantum Science and Technology (IQST), the German Research Foundation (DFG), the Federal Ministry of Education and Research (BMBF, project SiSiQ), and the Federal Ministry for Economic Affairs and Energy (BMWi, project PlanQK). AEJ is supported by the EPSRC Hub in Quantum Computing and Simulation (EP/T001062/1).

\end{footnotesize}

\newpage
\clearpage
\onecolumngrid
\appendix

\section{Calculating scattering probabilities}\label{appx:scatteringProbabilities}
If single photons with internal states $\rho_{i},i=1,\ldots,N$ are input into the $i$th arms of an $N$-mode interferometer, the probability of a detection outcome is given by \cite{Shchesnovich2015,Tichy2015}
\begin{equation}\label{eqn:mixedProbabilities}
P_{\rho_1,...,\rho_N}=\mathcal{N}\sum_{\sigma\in S_N}\bigg[\prod_{j=1}^{a}\text{Tr}(\rho_{\alpha_{j1}}\ldots\rho_{\alpha_{jn}})\bigg]
\times\text{perm}(M\star M^*_{\sigma,\mathds{1}}),
\end{equation}
where $(\alpha_{j1},..,\alpha_{jn})$ is the structure of the $j$th disjoint cycle of $\sigma$, where $\sigma$ is an element of the permutation group $S_{N}$. $a$ is the number of disjoint cycles in $\sigma$, $n$ is the length of the $j$th cycle, and $M$ is the scattering matrix constructed from the input and output mode occupations. Here $\star$ indicates the elementwise product of matrix elements. For an input state configuration with photon number occupations $n_i$, $r=(n_1,n_2,\ldots,n_i,\ldots,n_m)$ and a measured output state configuration $s=(l_1,l_2,\ldots,l_i,\ldots,l_m)$, mode assignment lists $d(r/s)$ are defined as
\begin{equation}
    d(r)=(\overbrace{1,..,1}^{n_1-\text{times}},\overbrace{2,..,2}^{n_2-\text{times}},...,\overbrace{m,...,m}^{n_m-\text{times}}),
\end{equation}
which contain the mode indices for each photon as many times as the number of photons occupying that mode. The scattering matrix $M$ is then constructed from the unitary $U$ as: $M=U_{d(r),d(s)}$. The normalisation $\mathcal{N}$ is given by: $\mathcal{N}=\left(\prod_{j}s_j!r_j!\right)^{-1}$.

\section{Insensitivity of coincident and fully bunched statistics to \texorpdfstring{$V_{abc}$}{V}}\label{appx:Vabc_insensitivity}
The coincident and fully bunched probabilities shown in Fig.~\ref{fig:fig4} for a tritter interferometer depend linearly on the pairwise traces and exhibit no dependence on the spanned volume $V_{abc}$ (defined for the pure state preparation in Fig.~\ref{fig:fig2}b,c), and this actually holds for any three-port unitary interferometer.

For scattering probabilities to exhibit sensitivity to the imaginary components of traces of density matrices, the expression in Eqn.~\ref{eqn:mixedProbabilities} should change under $\rho_{j}\rightarrow\rho_{j}^{*}$. $n$-photon interference depends on $\mathrm{Tr}\left(\rho_{\alpha_{j1}}\ldots \rho_{\alpha_{jn}}\right)$, where $(\alpha_{j1},...,\alpha_{jn})$ is the $j^{th}$ disjoint cycle of some permutation $\sigma$. Under complex conjugation of all $\rho_{j}$ we find:
\begin{equation}
    \left[\mathrm{Tr}\left(\rho_{\alpha_{j1}}\ldots \rho_{\alpha_{jn}}\right)\right]^{*}=\mathrm{Tr}\left(\rho_{\alpha_{jn}}\ldots \rho_{\alpha_{j1}}\right).
\end{equation}
The density matrices in the trace are now permuted according to the $j^{th}$ disjoint cycle of the inverse permutation $\sigma^{-1}$. $n$-photon interference in Eqn.~\ref{eqn:mixedProbabilities} therefore depends on:
\begin{equation}
    \text{Tr}(\rho_{\alpha_{j1}}\ldots\rho_{\alpha_{jn}})\times\text{perm}(M\star M^*_{\sigma,\mathds{1}})  +\text{Tr}(\rho_{\alpha_{j1}}\ldots\rho_{\alpha_{jn}})^{*}\times\text{perm}(M\star M^*_{\sigma^{-1},\mathds{1}}).
\end{equation}
For the case of three-photon coincidences from a three-port unitary interferometer $U$, the scattering matrix $M=U$. Inserting the Euler angle decomposition for a general $SU(3)$ unitary~\cite{Tilma_2002} reveals that these permanents are purely real. Hence coincidences here are not sensitive to the imaginary part of $\mathrm{Tr}(\rho_{a}\rho_{b}\rho_{c})$ and so exhibit no dependence on $V_{abc}$. For more than three photons this condition does not necessarily hold: $N$-photon coincidence probabilities, where $N>3$, can be sensitive to imaginary components of traces of density matrices.

For fully bunched probabilities where all $N$ photons occupy the $k$th output port, the scattering matrix $M$ is constructed by taking the $k$th column of the unitary describing the interferometer $N$ times. The elementwise product $M\star M_{\sigma,\mathds{I}}^{*}$ yields a real matrix and so a real permanent. Therefore fully bunched probabilities do not depend on imaginary parts of the traces of density matrices, and so in the three-photon case do not depend on $V_{abc}$.

\section{Experimental setup}\label{appx:setup}
A mode-locked Ti:Sapphire laser with a repetition rate of 76 MHz emits 3.2~ps pulses centred at 775~nm. It is used to pump two periodically poled potassium titanyl phosphate (ppKTP) crystals of dimensions 1 mm x 1 mm x 30 mm and a poling period of $\Lambda=46.175~\mu $m. The pump power is controlled by a combination of a HWP and a polarising beam splitter (PBS) before being divided into two beams using a 50:50 beam splitter. After additional power control, the beams are focused onto the crystals using a combination of lenses, resulting in an ideal beam waist at the crystals. HWPs before the crystals ensure the polarisation required for the phase-matching condition. The crystals are placed in temperature-controlled ovens to set the temperature for ideal degenerate phase-matching.

After the crystals,  longpass filters are used to block the pump beam; 3~nm and 1.5~nm bandpass filters centred at 1550~nm are used to ensure spectral indistinguishability, which is verified by measuring the photons' spectra using a spectrometer (Kymera 193i-B1 spectrograph and iDus InGaAs DU490A-1.7 photo-diode array by Andor Technology). The signal and idler photons are orthogonally polarised and are separated using PBSs. A linear polariser is used to clean the polarisation in the reflected arm of the PBS. HWPs and QWPs are then used to set the photons' polarisations. Delay stages temporally align the photons, which are then coupled into single-mode fibres. Paddles on the fibres allow compensation for random polarisation rotations before the photons interfere at a fused fibre-based tritter. In our setup, both sources were pumped with an average power of 50~mW, which is low enough to reduce the effect of higher-order emissions while maintaining good count rates. We obtain an average rate for four-photon events of $\approx 10 ~\text{Hz}$. This allows for shorter integration times per measurement, thereby reducing the effect of environmental changes.

\begin{figure}[ht]
	\centering
	\includegraphics[width=0.875\textwidth]{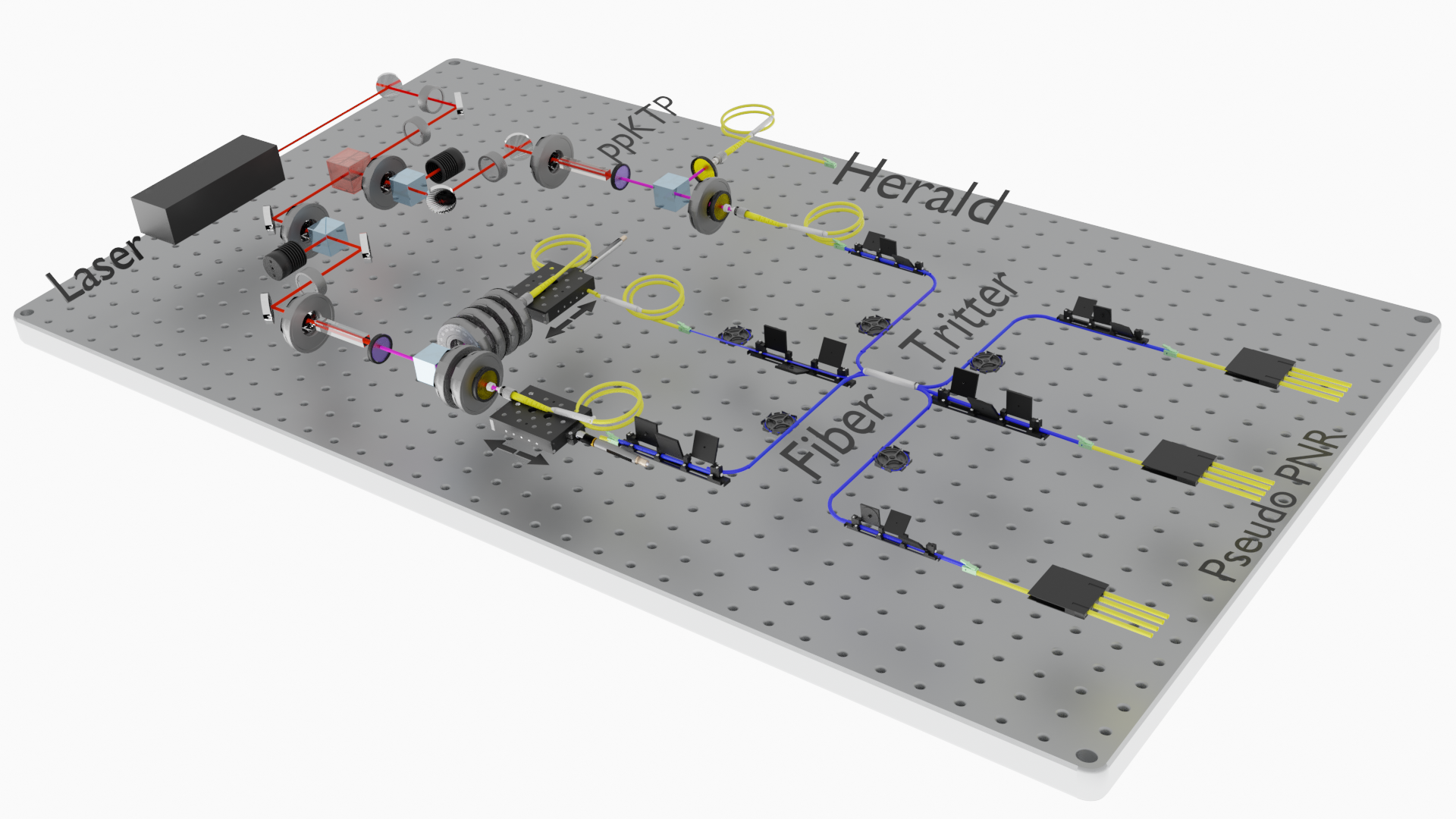}
	\includegraphics[width=0.7\textwidth]{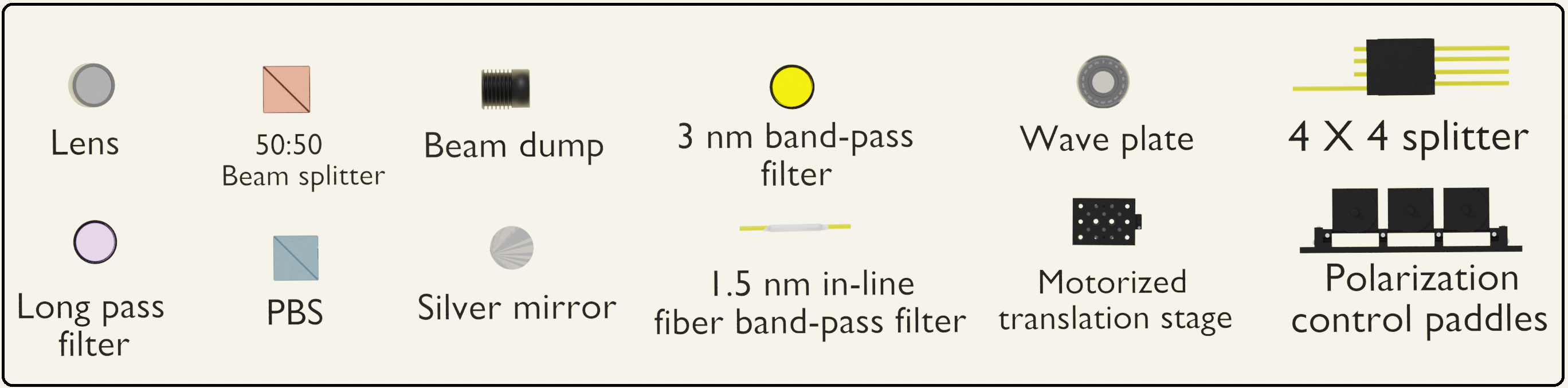}
	\caption{\label{fig:figsetupdetailed} Detailed experimental setup described in the main text of Appendix~\ref{appx:setup}.}
\end{figure}

\section{HOM dips and pairwise traces for three-photon experiments}\label{appx:HOMdata}
The two preparations of three photons presented in Eqns.~\ref{eqn:prep1} and~\ref{eqn:prep2} of the main text are chosen so that they cannot be discriminated by HOM visibilities alone. Alongside the three-photon data shown in Fig.~\ref{fig:fig4}, we also record sets of HOM dips and use their visibilities to infer the pairwise traces of pairs of interfering photons. Note that the two-photon coincidence probability through an ideal tritter is $P_{11}=(2-\TR{\rho_{j}\rho_{k}})/9$, and so here the maximum visibility of an indistinguishable HOM dip is $\mathcal{V}=0.5$.

\begin{figure*}[ht]
	\centering
	\includegraphics[width=1.\textwidth]{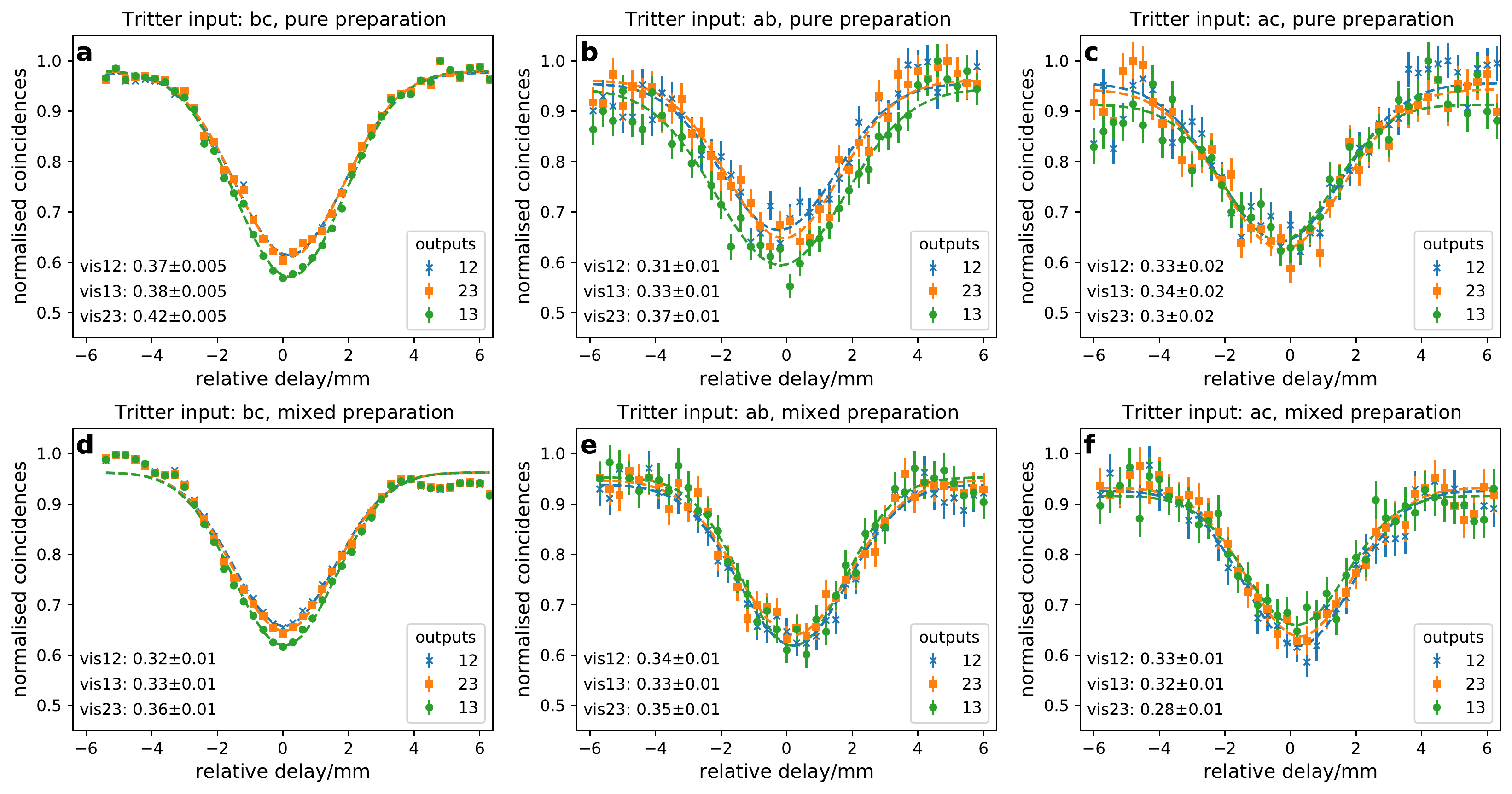}
	\caption{\label{fig:appx_HOM} Experimental HOM dips for pairs of photons from the pure (top row, \textbf{a}-\textbf{c}) and mixed (bottom row, \textbf{d}-\textbf{f}) preparations of photons, with pairwise traces set to $\TR{\rho_{j}\rho_{k}}=0.7$. Error bars are from Poissonian statistics (smaller for the single source twofolds in \textbf{a} and \textbf{d}), dashed lines are fits to the data, and quoted visibilities and errors are determined from these fits.}
\end{figure*}

As an example, we choose a pairwise trace of $\TR{\rho_{j}\rho_{k}}=0.7$. This corresponds to $\theta=0.684$ in the pure preparation of Eqn.~\ref{eqn:prep1} and a preparation probability $p=0.816$ in the mixed preparation of Eqn.~\ref{eqn:prep2}. In the latter case we simulate impurity by incoherently summing counts for the four combinations of input pairs of photons each in $H$ or $V$ polarisation. Results are shown in Fig.~\ref{fig:appx_HOM}. The key observation is that the HOM visibilities are the same for the pure preparation in the top row and the mixed preparation in the bottom row.

\begin{figure*}[ht]
	\centering
	\includegraphics[width=0.5\textwidth]{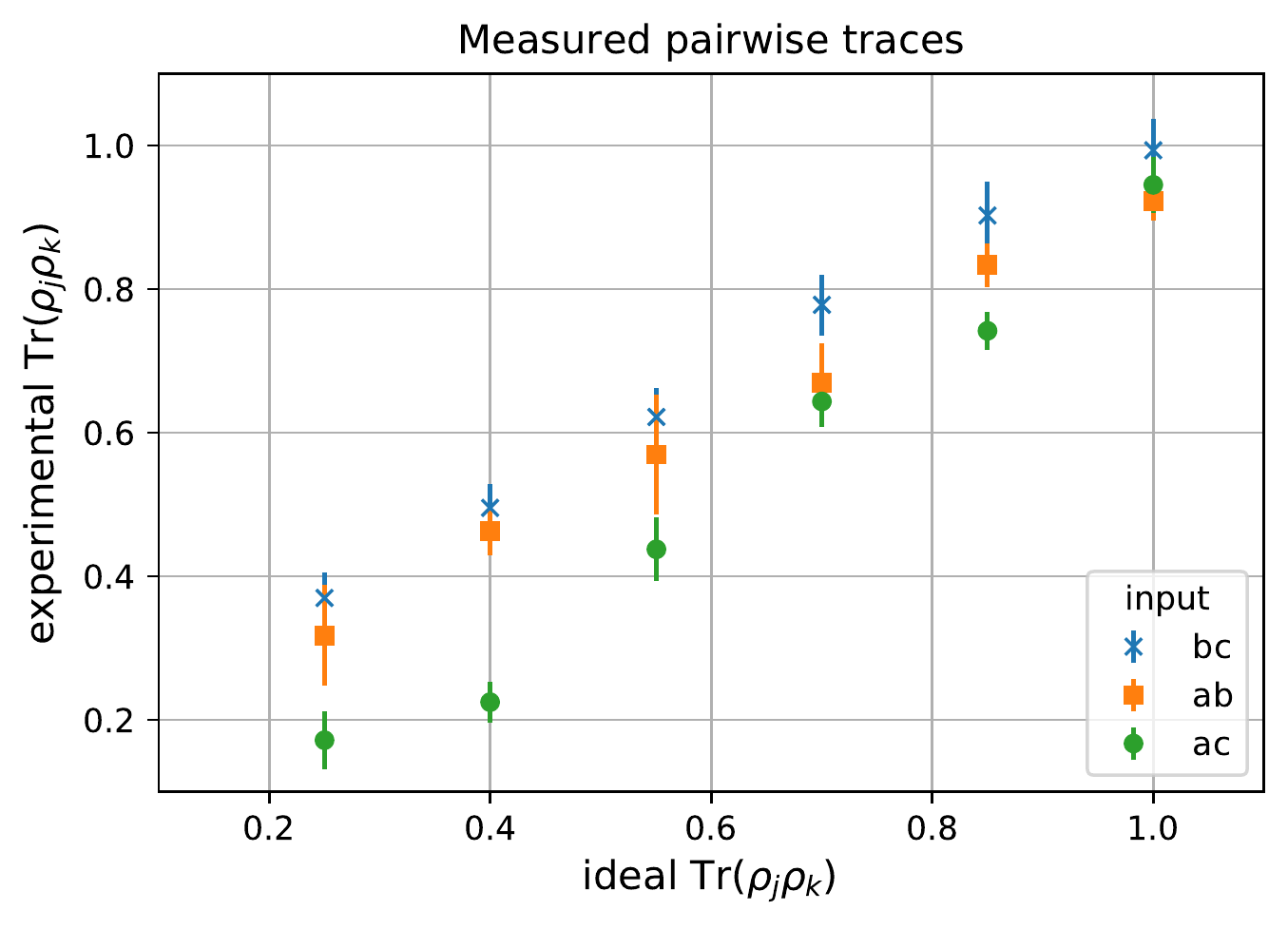}
	\caption{\label{fig:appx_pairwise} Plots of the experimentally determined pairwise traces for the pure state preparation of three photons, against the ideal value from the expression $\TR{\rho_{j}\rho_{k}}=(5+3\cos2\theta)/8$ for the states in Eqn.~\ref{eqn:prep1}. Error bars indicate the standard deviation on the estimate of the pairwise trace from the three output HOM dip visibilities.}
\end{figure*}

To confirm the accuracy of our pure state preparation, we fit all sets of HOM dips and plot the inferred pairwise traces against the ideal values. To a good approximation, we assume a perfect tritter so that the pairwise trace is given by half the associated HOM visibility. Results are shown in Fig.~\ref{fig:appx_pairwise}. Differences between ideal and experimental values are due to slight variations in the paths around the Bloch sphere depicted in Fig.~\ref{fig:fig2}b, likely arising from small errors in waveplate calibration. We measured HOM dip visibilities for six of the eleven $\theta$ values used in Fig.~\ref{fig:fig4}a-c. A linear fit is used to estimate pairwise traces that were not measured experimentally. The means of these pairwise traces and their errors are used for the x-coordinates and errors in the top row plots of Fig.~\ref{fig:fig4}.

\section{Using \texorpdfstring{$V_{abc}$}{V} to identify mixedness for qubits}\label{appx:identifyMixedness}
We can determine $V_{abc}$ using multiphoton statistics and here we show how it permits identification of mixedness. Its magnitude is given by~\cite{Casey1889}
\begin{equation}\label{eqn:Vabc}
\vert V_{abc}\vert=r_{a}r_{b}r_{c}\big[1-(\hat{\bm{r}}_{a}\cdot\hat{\bm{r}}_{b})^{2}-(\hat{\bm{r}}_{a}\cdot\hat{\bm{r}}_{c})^{2}-(\hat{\bm{r}}_{b}\cdot\hat{\bm{r}}_{c})^{2}
+2(\hat{\bm{r}}_{a}\cdot\hat{\bm{r}}_{b})(\hat{\bm{r}}_{a}\cdot\hat{\bm{r}}_{c})(\hat{\bm{r}}_{b}\cdot\hat{\bm{r}}_{c})\big]^{\frac{1}{2}},
\end{equation}
where $r_{i}$ is the length of vector $\bm{r}_{i}$ and $\hat{\bm{r}}_{i}=\bm{r}_{i}/r_{i}$ are unit vectors. For pure qubits $r_{i}=1$ and the magnitude $\vert V_{abc}\vert$ is completely determined by dot products of unit vectors, describing distinguishabilities that can be obtained from HOM visibilities (see Fig.~\ref{fig:fig1}c). The triple overlap of Eqn.~\ref{eqn:Trabc} reduces to $\braket{a}{b}\braket{b}{c}\braket{c}{a}$ and its argument, the triad phase $\varphi_{abc}$, is given by half the solid angle subtended by the three vectors and encodes three-particle distinguishability~\cite{Menssen2017}.

If the qubit states are not pure then $V_{abc}$ replaces $\varphi_{abc}$ as the appropriate collective distinguishing parameter and it can be used to identify mixedness in a way that is impossible using two-photon interference. 
HOM dips between partially distinguishable pairs of pure photons would yield $\vert V_{abc}\vert$ that satisfies Eqn.~\ref{eqn:Vabc} with all $r_{i}=1$. If this does not hold then the assumption of pure states is incorrect and mixedness can be identified. It is worth noting here why qubit internal states are insufficient to freely tune the five distinguishing parameters for three-photon interference: knowledge of the vector dot products fixes the magnitude $\vert V_{abc}\vert$ and so the real and imaginary parts of $\TR{\rho_{a}\rho_{b}\rho_{c}}$ are not independent.

We now briefly describe an experiment we performed where the measured $V_{abc}$ indicates mixedness of an internal state. We prepare three photons in pure polarisation states labelled $\ket{a},\ket{b},\ket{c}$ (with associated Bloch vectors $\bm{r}_{j},j=a,b,c$) that ideally set the following quantities:
\begin{equation}
    \begin{aligned}
        \bm{r}_{a}\cdot\bm{r}_{b} &= 0.5,\\
        \bm{r}_{a}\cdot\bm{r}_{c} &= 0.27,\\
        \bm{r}_{b}\cdot\bm{r}_{c} &= -0.03,\\
        V_{abc} = \bm{r}_{a}\cdot(\bm{r}_{b}\times\bm{r}_{c}) &= -0.82.
    \end{aligned}
\end{equation}
We then perform two sets of measurements: HOM dip visibilities between pairs of photons to infer experimental Bloch vector dot products (from pairwise traces, see Eqn.~\ref{eqn:TrAB}), and three-photon counts at the tritter outputs as the temporal delay of photon $b$ is swept: this varies between temporal distinguishability and indistinguishability of $b$ with respect to the other photons. The relative values of the partially bunched probabilities at these two extremes allows direct measurement of $V_{abc}$ (using Eqn.~\ref{eqn:mixedDistProbs}).

In order to simulate mixedness of the internal state $\rho_{a}$, we repeat the above measurements but now with the first photon in the pure state $\ket{a^{\perp}}$ which is on the opposite of the Bloch sphere to $\ket{a}$. This flips the sign of $V_{abc}$ above and also changes some of the dot products. We then take weighted sums of statistics for the first and second preparations of the photons to simulate state impurity.

As an example, we set the purity of the first photon to 0.9 so that the associated Bloch vector length is ideally $r_{a}=0.64$. Substituting measured values of vector dot products and $\vert V_{abc}\vert$ into Eqn.~\ref{eqn:Vabc}, we experimentally find a best estimate $r_{a}=0.56$. This confirms that measurement of $V_{abc}$ can permit identification of mixedness of a qubit internal state. Deviation from the ideal value mostly arises from imperfect state preparation, fitting errors due to Poissonian counting statistics and low visibility signals, and residual spectral distinguishability.

\section{Trace of four qubit internal states}\label{appx:fourQubits}
The interference of four photons will depend on pairwise traces, triple traces, and also the quadruple trace $\TR{\rho_{a}\rho_{b}\rho_{c}\rho_{d}}$. For qubit internal states, this last quantity can be expressed in terms of Bloch vector dot and scalar triple products:
\begin{equation}
\begin{aligned}
    \TR{\rho_{a}\rho_{b}\rho_{c}\rho_{d}}&=\frac{1}{8}\bigg[1+[\bm{r}_{a}\cdot\bm{r}_{b}+\bm{r}_{a}\cdot\bm{r}_{c}+\bm{r}_{a}\cdot\bm{r}_{d}+\bm{r}_{b}\cdot\bm{r}_{c}+\bm{r}_{b}\cdot\bm{r}_{d}+\bm{r}_{c}\cdot\bm{r}_{d}]\\
    &+[(\bm{r}_{a}\cdot\bm{r}_{b})(\bm{r}_{c}\cdot\bm{r}_{d})-(\bm{r}_{a}\cdot\bm{r}_{c})(\bm{r}_{b}\cdot\bm{r}_{d})+(\bm{r}_{a}\cdot\bm{r}_{d})(\bm{r}_{b}\cdot\bm{r}_{c})]\\
    &+i[\bm{r}_{a}\cdot(\bm{r}_{b}\times\bm{r}_{c})+\bm{r}_{a}\cdot(\bm{r}_{b}\times\bm{r}_{d})+\bm{r}_{a}\cdot(\bm{r}_{c}\times\bm{r}_{d})+\bm{r}_{b}\cdot(\bm{r}_{c}\times\bm{r}_{d})]\bigg].
\end{aligned}
\end{equation}
Hence two- and three-photon parameters fully determine four-photon interference for mixed qubit internal states. For qubits, higher-order traces can be decomposed into combinations of dot and scalar triple products using the commutation relations for Pauli matrices: $\sigma_{j}\sigma_{k}=\delta_{jk}\mathds{I}+i\varepsilon_{jkl}\sigma_{l}$.

\section{Interference of three photons with qutrit internal states}\label{appx:qutrits}
In the main text we concentrated on three photons with qubit internal states to present an intuitive geometric picture of mixedness and distinguishability. However, this imposes a restriction on the five parameters governing three-photon interference through Eqns.~\ref{eqn:Trabc} and ~\ref{eqn:Vabc}, and as discussed in Appendix~\ref{appx:identifyMixedness}. A three-dimensional (qutrit) internal space is needed to fully probe three-photon distinguishability.

The Gell-Mann matrices $\{\lambda_{i}\}$ are generators of the SU(3) group and satisfy commutation and anticommutation relations
\begin{equation}
    [\lambda_{r},\lambda_{s}]=2if_{rst}\lambda_{t}, \quad \{\lambda_{s},\lambda_{r}\}=\frac{4}{3}\delta_{rs}+2d_{rst}\lambda_{t}.
\end{equation}
$d_{rst}$ and $f_{rst}$ are respectively the completely symmetric and antisymmetric $SU(3)$ structure constants. These define symmetric and antisymmetric vector products~\cite{Arvind_1997}:
\begin{align}
    \bm{a}\star\bm{b}&\coloneqq\sqrt{3}d_{rst}a_{s}b_{t},\quad \bm{a}\star\bm{b}=\bm{b}\star\bm{a},\\
    \bm{a}\wedge\bm{b}&\coloneqq f_{rst}a_{s}b_{t},\quad \bm{a}\wedge\bm{b}=-\bm{b}\wedge\bm{a}.
\end{align}
$\bm{a}$ and $\bm{b}$ are eight-dimensional real vectors that are the qutrit equivalents of Bloch vectors.

A general qutrit state $\rho_{j}$ can be expressed by~\cite{Arvind1997}
\begin{equation}
\rho_{j}=\frac{1}{3}\left(\mathbb{I}+\sqrt{3}\bm{n}_{j}\cdot\boldsymbol{\lambda}\right),
\end{equation}
where $\boldsymbol{\lambda}$ is the vector of the eight Gell-Mann matrices and the components of the eight-dimensional vector $\bm{n}_{j}=\sqrt{3}/2\times\TR{\rho_{j}\boldsymbol{\lambda}}$. Pairwise traces of two qutrit states are, as for qubits, captured by the dot products of the associated vectors. However, the triple trace becomes considerably more complicated~\cite{Arvind1997,Hartley2004,Caves2000}:
\begin{equation}
    \TR{\rho_{a}\rho_{b}\rho_{c}}=\frac{1}{9}\left[\left(1+2(\bm{n}_{a}\cdot\bm{n}_{b}+\bm{n}_{a}\cdot\bm{n}_{c}+\bm{n}_{b}\cdot\bm{n}_{c}+\bm{n}_{a}\cdot(\bm{n}_{b}\star\bm{n}_{c})\right)+i\frac{2}{3\sqrt{3}}\bm{n}_{a}\cdot(\bm{n}_{b}\wedge\bm{n}_{c})\right].
\end{equation}
For qubits we saw from Eqn.~\ref{eqn:Trabc} that the real part of the triple trace is fully determined by Bloch vector dot products. For qutrits this no longer holds due to the presence of $\bm{n}_{a}\cdot(\bm{n}_{b}\star\bm{n}_{c})$. It is also possible to prepare three pure qutrit states such that the pairwise traces are constant but the terms involving three $\bm{n}_{j}$ vectors vary -- see for example Ref.~\cite{Menssen2017} or the following configuration from~\cite{Hartley2004}:
\begin{equation}
    \begin{aligned}
    \ket{a}&=\ket{\bm{0}},\\
    \ket{b}&=\frac{1}{\sqrt{2}}\left(\ket{\bm{0}}+\ket{\bm{1}}\right),\\
    \ket{c}&=\frac{1}{\sqrt{3}}\left(\ket{\bm{0}}+(2e^{i\gamma}-1)\ket{\bm{1}}+\sqrt{4\cos\gamma-3}\ket{\bm{2}}\right).
    \end{aligned}
\end{equation}
The sum of the dot products is $3/4$, independent of the angle $\gamma$, but the triple trace is $e^{i\gamma}/3$.

\end{document}